\newcommand{\co}{ {\cal O}}
\newcommand{\beq}{ \begin{equation}}
\newcommand{\eeq}{ \end{equation}}
\newcommand{\beqau}{\begin{eqnarray*}}
\newcommand{\eeqau}{\end{eqnarray*}}
\newcommand{\beqa}{\begin{eqnarray}}
\newcommand{\eeqa}{\end{eqnarray}}
\newcommand{\fig}[1]{Figure~\ref{#1}}
\newcommand{\tab}[1]{Table~\ref{#1}}

\newcommand{\half}{\frac{1}{2}}

\newcommand{\ket}[1]{| {#1} \rangle}

\newcommand{\fltfig}[3]{
\begin{figure}[htbp]
\begin{center} 
#2
{\large \caption{ \label{#3} #1 } }
\end{center}
\end{figure}
}
\newcommand{\flttab}[3]{
\begin{table}[htbp]
{\large \caption{ \label{#3} #1 }}
\begin{center} 
#2
\end{center}
\end{table}
}
\newcommand{\epsfaxhax}[2]{
        \centerline{
          \hspace{-20pt}
          \epsfxsize=225pt
          {\epsfbox{#1}}
          \hspace{-15pt}
          \epsfxsize=225pt
          {\epsfbox{#2}}}
}
\newcommand{\epsfaxhaxhax}[3]{
        \centerline{
          \hspace{-20pt}
          \epsfxsize=160pt
          {\epsfbox{#1}}
          \hspace{-15pt}
          \epsfxsize=160pt
          {\epsfbox{#2}}
          \hspace{-10pt}
          \epsfxsize=160pt
          {\epsfbox{#3}}}
}
\newcommand{\pspicture}[1]{\centerline{\setlength\epsfxsize{225pt}\epsfbox{#1}}}

\documentstyle[aps,tighten,epsf]{revtex}
\begin{document}
\preprint{
\begin{minipage}{4.5cm}Edinburgh Preprint 99/3\\
GUTPA-99-02-02\\\end{minipage}}

\draft
\author{P. Boyle
\thanks{Present Address: Department of Physics and Astronomy, 
University of Glasgow, Glasgow}}
\address{Department of Physics and Astronomy,\\University of Edinburgh,\\Edinburgh.}
\author{UKQCD Collaboration}
\date{\today}
\title{A Novel Gauge Invariant Multi-State Smearing Technique}
\maketitle
\begin{abstract}
We present an investigation of a gauge invariant smearing
technique that allows the construction of smearing functions
with arbitrary radial behaviour, by foresaking the space filling
nature of traditional smearing techniques. This is applied to
both heavy-heavy, heavy-light, and light-light systems with one
particular choice of radial ``wavefunction'' - the hydrogenic
solutions - and we find good stability for both fitted masses
and amplitudes of the radially excited states. The dependence of
the amplitudes on the smearing radius is demonstrated to be well understood,
while near optimal smearing radii may be found with extremely low statistics
using a property of the smeared-local correlator.
The smearing technique is inexpensive since it is non-iterative, achieves
a good signal to noise ratio, and can be altered to use
wavefunctions from, say, potential models or the Bethe-Salpether equations
in future simulations.
\end{abstract}
\pacs{}

\section{Introduction}

{\it Smearing} is a technique that has been used for some time in 
Lattice QCD simulations \cite{wuppertal_sm,jacobi_sm,cmi_fuzz} in order to 
improve the quality of the results
obtained for a given computational expense by reducing the extent
of excited state contamination. The choice of operator used to represent a
state in any given simulation is essentially free within the constraint that
it must have the appropriate quantum numbers of the desired state, which (usually)
guarantees non-zero overlap with the groundstate of those quantum numbers. 
In Euclidean space, all such operators asymptotically behave as the 
lowest lying state with quantum numbers matching those of the operator 
\footnote{Ignoring questions
such as scalar glueball mixing in quenched simulations, and as we shall see
special cases where the overlap is fine tuned to be near zero.}.
We consider the two point correlation function of a set of operators $\co_k$ with 
some definite $J^{PC}$ in a Euclidean lattice of temporal extent $T$.
Inserting a complete set of physical intermediate states of the same quantum numbers we 
find the same asymptotic behaviour for all operators
\begin{eqnarray}
\langle \co_k(x) \co_k^\dagger(0)\rangle &=& \sum\limits_n 
 \frac{\langle 0|\co_k|n\rangle \langle n|\co_k^\dagger|0\rangle}{2E_n}
 e^{-E_n\frac{T}{2}}\cosh E_n(t-\frac{T}{2})\\
&=& \sum\limits_n C_{kn} C_{kn}^\ast 
\frac{ e^{-E_n\frac{T}{2}}}{2 E_n} \cosh E_n(t-\frac{T}{2})\\
&\stackrel{t\to\infty}{\longrightarrow}& |C_{k0}|^2 \frac{ e^{-E_n\frac{T}{2}}}{2 E_0} \cosh E_0(t-\frac{T}{2})
\end{eqnarray}
This somewhat egalitarian situation is the principal reason why lattice calculations have
traditionally had difficulty extending their scope to radially excited states. 
However, some choices of operators turn out to be more equal than others, depending on the
relative values of the $C_{nk}$. For ground state phenomenology we would like
$C_{0k} \gg C_{nk} ~~~ \forall n \geq 1$. To this end it is common to construct extended 
operators which project more predominantly onto the groundstate
of the desired system; it is this extension procedure that has
been dubbed smearing. Typically we require that the smearing procedure be parity
and charge positive, and possess at least cubic symmetry

There are two different schools of thought on how to construct smeared operators  - 
gauge invariant smearing and gauge fixed smearing.

\subsection{Gauge Invariant Smearing}
Explicit gauge invariance is one of the most
attractive features of non-perturbative simulation techniques, and remains
to this day a necessary (but sadly insufficient) condition for correct operation
of simulation code. Further the gauge fixing process may be multiply defined 
due to Gribov ambiguities of some gauge fixing conditions \cite{Gribov,Parrinello},
and one has to assume that the ambiguity does not cause any bias in the quantities
being measured. For these reason many people are unwilling to use gauge fixing, and proceed
with one of a number of gauge invariant techniques.

The most commonly used gauge invariant techniques are Wuppertal \cite{wuppertal_sm}, Jacobi
\cite{jacobi_sm}, and fuzzed smearing \cite{cmi_fuzz}. 
Wuppertal smearing covariantly smears the quark fields with a spatial distribution which
is the solution of the scalar Klein Gordon equation on a given gauge configuration, while
Jacobi smearing is a numerically efficient way to approximate this.
In order to preserve
gauge invariance both the first two methods iteratively apply a
gauge invariant nearest neighbour operator to smear out a quark field, with limiting 
behaviour to weight sites in a ``bell shaped'' distribution. 
Given the extended nature of the physical states, and the wavefunction picture 
in the non-relativistic limit one would expect smeared operators to more 
accurately project onto the ground state, and indeed this is borne out by simulations.
However these gauge invariant techniques give little control over the
precise functional form of the smearing beyond a basic radius parameter.

\label{fuzzingsection}

The cost of iteratively applying, say, the Jacobi operator $N$ times on a propagator
is somewhat prohibitive. Lacock and Michael \cite{cmi_fuzz} suggested
the use of a covariantly transported ``cross'' using low noise \emph{fuzzed}
links.
The fuzzing prescription iterates
\beq
U_j^\prime( x) = c~U_j(x) +  \sum\limits_{i\neq j}\left[ U^{\rm Staple}_i( x,x+\hat{j})+ 
U^{\rm Staple}_{-i}( x,x+\hat{j}) \right],
\eeq
\beq
U_j(x) = {\cal P} ~U_j^\prime(x) ,
\eeq
where ${\cal P}$ is a projection onto SU(3) (not closed under addition) 
by Cabibbo Marinari maximisation of ${\rm Tr} [U_j^\dagger U_j^\prime]$ with six 
hits on each SU(2) subgroup.
Typically $c$ is set to 2.0, and there are 5 iterations performed.
These fuzzed links are used to transport the quark field by some number
of sites $N$ along each of the spatial axes.
\beq
\psi^\prime(x) = \sum_{i=x,y,z} 
\left[ \prod_{n=1}^r U_i(x+(n-1)\hat{i})\right]\times \psi(x+r\hat{i})
+ \left[ \prod_{n=1}^r U_i^\dagger(x-n\hat{i})\right]\times \psi(x-r\hat{i})
\eeq
The fuzzing radius, $r$, is chosen to minimise the contamination of 
the ground state operator.

\subsection{Gauge Fixed Smearing}

Gauge fixed smearing (either Coulomb or Landau gauge) allows complete freedom for the functional
form of the smearing; in a fixed gauge we are freed from the need to iterate 
local covariant transport to space fill, and convolve quark fields with arbitrary
smearing functions using fast Fourier transforms. In particular, the use of 
smearing functions with nodes has allowed reliable extraction of excited 
state energies from NRQCD through the use of multi-exponential fits \cite{Upsilon}.
Neither the cost of Fourier transforming the data set multiple times to implement this smearing
on a massively parallel machine nor the loss of gauge invariance are desirable features,
and there is clearly a need for a gauge invariant smearing technique with free radial
functional form.

In order to proceed, we consider the non-relativistic interpretation of $q\bar{q}$ systems
in terms of spatial wavefunctions. This is certainly appropriate to heavy-heavy mesons,
however, we can apply this technique in heavy-light and light-light 
systems without needing to worry about its relevance.

\section{Smearing In A Non-relativistic Potential}
We consider smearing in the context of a non-relativistic potential model
with spherical symmetry. 
The solutions may be written as $\psi_{nlm}(r,\theta,\phi) = R_{nl}(r) P_{lm}(\theta,\phi)$
where $ P_{lm}(\theta,\phi)$ are the usual spherical harmonics. The orthogonality relation
of the wavefunctions is then
\begin{equation}
\int dV \psi_{mko}^*\psi_{nlp} =
\delta_{kl}\delta_{op} \int\limits_{r=0}^{\infty}R_{ml}^*(r) R_{nl}(r)r^2 dr 
\end{equation}
\begin{equation}
\Rightarrow \int\limits_{r=0}^{\infty}  R_{ml}^*(r) R_{nl}(r)r^2dr = \delta_{mn}
\end{equation}
We shall ignore the $l$ and $m$ indices from here, and consider spherically symmetric
functions.
Suppose we construct  non-local meson operators

\beq
{\cal O}_n(x) = \int d^3y \bar{\psi}(y) S_n(y,x) \psi(x)
\eeq
where the smearing function $S_n(y,x)=S_n(y-x)$ is thought of as a spatial wavefunction
defined on lattice sites
\beq
S_n(y,0) \equiv S_n(y) \sum_{i,j,k} \delta^3 ( y - i \hat{x} - j \hat{y} - k \hat{z} )
\eeq
Note we have dropped the colour indices since in our wavefunction
approximation the gauge degrees of freedom are represented by 
a colour singlet potential.

The operator ${\cal O}$ will have an overlap with each physical state
$\ket{\psi_m}$ of the  relevant $J^{PC}$
\beq
\label{discrete_smearing_integral}
C_{nm} = \int \psi_m^* (x) S_n(x) dV = \sum_{i,j,k}
\psi_m^* (i \hat{x} + j \hat{y} + k \hat{z}) S_n(i \hat{x} + j \hat{y} + k \hat{z} )
\eeq
Now if $S_n(x)$ is chosen to well approximate the state $\psi_n(x)$ and 
the lattice is sufficiently fine that the sum (\ref{discrete_smearing_integral})
is close to the integral 
\beq
C_{nm} = \int \psi_m^* S_n(x) dV \simeq \int \psi_m^*(x) \psi_n(x) dV = \delta_{nm}
\eeq
Thus with a well chosen set smearing functions $S_n$ we can write
\beq
C_{nm} = \delta_{nm} + \eta_{nm}
\eeq
where $\eta_{nm}$ is a small contamination to the signal, giving a set
of correlation functions corresponding to each of the physical 
radial excitations. In fact $\eta_{nm} \simeq 0 ~~\forall m < n$
guarantees the lowest state significantly
contributing to the correlation function of the
$n$th radially excited operator is
in fact the $n$th state, and the effective mass will not decay to the
ground state of the $J^{PC}$ on a finite lattice.

If we choose a non-space-filling smearing function of the form
\beq
S_n(\vec{x},0) \equiv  \sum_{r=0}^{N} r^2 \phi_n(r) \sum_{\hat{\mu}=\hat{x},\hat{y},\hat{z}}
\left( \delta^3( \vec{x} - r\hat{\mu} ) + \delta^3(\vec{x} + r\hat{\mu} )\right)
\eeq
where $\phi_n$ is some arbitrary modulating function that is chosen to
approximate the true solution $\psi_n$, and the factor $r^2$ compensates for the
non-space-filling nature of the smearing,
then the overlap is
\beqa
C_{nm} &=& \int \psi_m^* (\vec{x}) S_n(\vec{x},0) dV \\
&=& \sum_{r=0}^{N} r^2 \phi_n(r)\psi_m^* (r)  \\
&\stackrel{a\to0}{\to}& \int\limits_{r=0}^\infty r^2 \psi_m^*\phi_n dr\\
&\simeq&\delta_{nm}.
\eeqa
Thus if we choose the $\phi_m$ correctly (i.e. to be $\psi_n$) then we establish
the orthogonality relation required without the need for space filling.
The new smearing technique proposed is a generalisation of the 
above form to incorporate colour:
\beq
\psi(x) \rightarrow  
\sum\limits_{r=0}^{N} {\scriptstyle (r+\half)}^2 \phi_n({\scriptstyle r}) 
 \sum\limits_{i=x,y,z} \left\{
\left[ \prod\limits_{n=1}^r U_i({\scriptstyle x+(n-1)\hat{i}})\right]
\psi({\scriptstyle x+r\hat{i}})
+ \left[ \prod\limits_{n=1}^r U_i^\dagger({\scriptstyle x-n\hat{i}})\right]
\psi({\scriptstyle x-r\hat{i}})
\right\}
\eeq
where the links are fuzzed links, c.f. Section \ref{fuzzingsection}.
We use the factor $(r+\frac{1}{2})^2$ as opposed to $r^2$ so that there is a 
non-zero contribution from the local current, since this was observed to improve
the statistical noise.
The procedure is very similar in cost to fuzzing with a radius $N$, since all the
terms can be formed as part of the process of fuzzing to the largest radius.
The technique therefore eliminates the two main problems with implementing 
gauge-invariant multi-state smearing, namely that gauge invariance
requires expensive iterative space filling techniques, and that it 
restricts the functional form.
Here we have a low-cost gauge invariant technique with an arbitrary functional
form, allowing the insertion of both  ground and radially excited wavefunctions.

We shall consider the case when the physical solutions are approximately
hydrogenic wavefunctions.

\section{Optimising the Smearing}

We consider the correlation function with a smeared source denoted $R$ and
a local sink denoted $L$, and restrict the argument to one 
contaminating excited state, and ignore the periodicity
of the lattice
\beq
C_t = A e^{-E_{1} t} + B e^{-E_{2}t}
\eeq
where
$
A \propto C_{L1}C^\ast_{R1}
$ 
and 
$
B \propto C_{L2}C^\ast_{R2}
$. 
The effective mass of this correlator is loosely the negative of the
derivative of the log:
\beq
M_t \simeq - \frac{d}{dt} \log C_t = \frac{E_1 + \frac{B}{A} E_2  e^{- (E_2-E_1) t}}
{1 + \frac{B}{A} e^{-(E_2 - E_1) t}} .
\eeq
By inspection it can be seen that there are two distinct cases for the approach to
the plateau:

\begin{itemize}
\item{
$\frac{B}{A} > 0$  ~~ approach from above
}
\item{
$\frac{B}{A} < 0$  ~~ approach from below, vertical asymptote at $t= \frac{\log |\frac{B}{A}|}{E
_2-E_1}$
}
\end{itemize}

\label{SmearOptTheory}
Suppose the physical wave functions are well described by the hydrogenic wavefunctions, $\psi_n$,
with some Bohr radius, $r_1$. We wish to approximate these solutions with a trial
wavefunction, which we choose to be the hydrogenic solutions $\phi_m$ with Bohr radius $r_2$.
Consider the overlap of the trial excited state with the physical groundstate,
\beqa
C_{10} &=& \int r^2 \psi_0\phi_1 dr \\ 
&\propto& \int r^2 (1-\frac{r}{2r_2}) e^{-\frac{r}{2r_2}}e^{-\frac{r}{r_1}} dr\\
&=& c^3\left[ \Gamma(3) - \frac{c}{2r_2} \Gamma(4) \right]\\
&\simeq& 2 c^3 \frac{\delta}{r_2}
\eeqa
where $c=\frac{2r_1r_2}{r_1+2r_2}$, ~$\delta = r_2 - r_1$, and the 
last step is made for small $\delta$.
Thus the overlap of the excited state smearing with the physical state 
is proportional to the 
error in the Bohr radius. The overlap is positive 
for smearing radii that are too large, and 
negative for radii that are too small, and so
the effective mass of the smeared - local correlation function
will have a vertical asymptote for radii that are too small, 
and a smooth form for radii that are too large.
This dramatic
change of behaviour may be used to tune the Bohr radius on very small statistical samples.

In fact it should always be possible
to orthogonalise the first radially excited 
smearing with respect to the physical groundstate
by tuning the Bohr radius parameter even when the physical states are
not hydrogenic. 
One choice for the radius will not, however, simultaneously minimise the overlaps
$C_{10}$ and $C_{01}$. 
The better the ansatz for the smearing functions, the closer together we expect the optimal 
radii for the groundstate and excited state correlators to lie. Where the ansatz
is better we also expect there to be less contamination from the higher excited states, since
the physical wavefunctions are all mutually orthogonal.

In order to demonstrate the applicability of this smearing technique, 
a feasibility study was carried out on a small sample, with a large number of smearing radii, 
in the light-light, heavy-light and heavy-heavy sectors.

\section{Feasibility Study }
We use the notation $R_{nl}$ to denote smearing with the corresponding
hydrogenic wavefunction with some Bohr radius.
Fourteen quenched configurations of a $24^3\times48$ lattice at $\beta=6.2$ were used 
with non-perturbatively $O(a)$ improved quarks (i.e. $C_{sw} = 1.61377$). Kappa values
0.13460 and 0.12300 were used, corresponding to quarks near the strange 
and charm masses respectively.
The lighter quark propagators had local sources, and both local and $R_{10}$
sources were generated for the heavy propagator
with Bohr radii of 2.4, 2.5, 2.6, 2.75, 2.8, 2.9, 3.0, 3.1, 3.25, 4.0, and 5.5
in lattice units. This allowed the smearing combinations
in \tab{TabFeasSmearingComb}
to be generated for each Bohr radius. The correlation functions were analysed using
uncorrelated fits, because the small statistical sample caused the correlation matrix to 
be too noisy for reliable inversion in correlated fits. As such the estimates of 
$\chi^2$ are probably unreliable.

Of primary interest is the simultaneous
two exponential fit to the $R_{10}$ and $R_{20}$
source smeared - local sink correlation functions. 
We show the fitted effective masses for various 
Bohr radius for the heavy-heavy, heavy-light and light-light 
sectors in Figures \ref{FIGBOYLE1} through \ref{FIGBOYLE11}.

The effective mass for the radially excited smearing demonstrates the 
expected transition in behaviour,
and appears to be forming a plateau for 
$r_0 = 2.6 (3.0 , 3.1)$ in the heavy-heavy (heavy-light, light-light)
sector. While the formation of an excited state plateau is clearly strongly
dependent on the selection of $r_0$ to within about $0.1$ (this is possible to do
using only one configuration by looking for vertical asymptote in the effective mass plot), 
we do not wish the fitted values to be dependent on the smearing.
To investigate this we study the stability of the fitted masses with the input parameters
to the fitting procedure, namely the Bohr radius, and the timeslice range of the fit.

\subsection{Stability Analysis of Fitted Masses}
The fitted values for the ground state and excited state masses are given in 
\tab{TABHHStab} to \tab{TABLLStab}, and are
found to be remarkably stable in the
fit range typically agreeing within statistical
error for all values of $t_{min} \ge 3$.

The level of statistical error on the excited state mass is also particularly pleasing.
The error on the excited light-light state is typically between only 5 and 10 times
larger than that on the ground state.

In order to study the stability of the fits with respect to the chosen Bohr radius,
I tabulate the fitted mass values corresponding to $t_{min}=4$ for all the simulated
Bohr radii in \tab{TabHHBohrStab}, \tab{TabHLBohrStab}, and \tab{TabLLBohrStab}.
They are stable within the statistical error
over almost the entire range of radii simulated.

\subsection{Selecting the Bohr Radius}
From inspection of the effective mass plots presented, we can see that the
optimal values of the Bohr radius parameter (at least for the $R_{20}$ correlator)
were about 2.6 for the heavy-heavy state, 3.0 for the heavy-light and 3.1 for the
light-light. Using the discussion in Section \ref{SmearOptTheory} we plot the fitted amplitudes
of the local-smeared correlation functions versus Bohr radius, looking for the zero
of the overlap of the $R_{20}$ correlator with the groundstate.

\fig{FigHHAmps}, \fig{FigHLAmps} and \fig{FigLLAmps} present the dependence 
of the four fitted amplitudes on the Bohr radius used in the heavy-heavy, heavy-light
and light-light systems respectively.
The linear fits are a good approximation, producing the optimal values for
the smearing radii given in \tab{TabOptimalRadii}.

While the effective mass plots presented were on fourteen configurations, it is clear that
following the pole on the plot gave very accurate optimisation of the Bohr radius,
as demonstrated by later fits to the amplitudes. 
It was anticipated in Section \ref{SmearOptTheory} that, since our smearing function basis
is  not the physical one, the radius for
minimal contamination of the groundstate by the excited state would
differ slightly from the radii which gave minimal contamination of the excited state.
This is not apparent from the data presented in \fig{FigHHAmps}, \fig{FigHLAmps} and \fig{FigLLAmps},
where there is no minimum of the $R_{10}$ - excited state amplitude. We must bear in mind 
however that the contamination
of the $R_{10}$ correlator is measured at early times, 
whereas the contamination of the $R_{20}$ correlator
is found from its asymptotic time dependence, leaving the determination of the optimal Bohr
radius from the  $R_{10}$ correlator much more subject to contamination from 
higher excited states.
For this reason, it is dangerous to look only at the local-smeared amplitude, since cancelling
contributions from the higher states can cause a seemingly flat effective mass. 

Instead we study
the smeared-smeared correlators, which have positive definite contaminating contributions,
and minimise the amplitude for the contamination of the $R_{10}$ - $R_{10}$ correlator. These
correlation functions were only generated for the heavy-heavy combination, and 
\fig{FigSScontamination} presents a sample fit to the pseudoscalar data.
Here the contamination below timeslice 5 is apparent, as well as a weaker drift
to the plateau beyond this point. It is this ultimate drift we must study.
\fig{FigSSamp} plots the fitted contaminating amplitude for this correlator versus radius;
we expect from Section \ref{SmearOptTheory} a quadratic dependence on the smeared-smeared
correlator. We obtain a minimising radius of $r_0 = 2.1(4)$ with a quadratic fit.

\subsection{Use of Doubly Smeared Correlation Functions}

In this section I demonstrate that the signal obtained from doubly smeared correlators
is useful. The double smearing is created by constructing the meson propagator from
two propagators which are each source smeared, while, for the purposes
of this demonstrate, we use local sinks.
The double exponential fit to
the double $R_{10}$ and single $R_{20}$ smeared correlation functions are presented
in \fig{FigHHDsmear} and \fig{FigLLDsmear} for the heavy-heavy and light-light systems 
respectively.

Clearly this smearing combination, while not tailored for any particular state constitutes
a valid signal that can be included in multi-channel multi-exponential fits, which is always of 
use, especially for the p-states.

\section{Comparison With Fuzzing}

In order to fairly compare the level of statistical error with traditional
techniques, I present double exponential fits to the $R_{10}$ sink and the fuzzed
sink, both with the optimal radius in \fig{FigHHfuzzComp},
\fig{FigHLfuzzComp} and \fig{FigLLfuzzComp} for the heavy-heavy,
heavy-light and light-light systems respectively, on the same set of
local source propagators. 

It can be seen that the 
statistical error from the $R_{10}$ correlator is much reduced with respect to
the fuzzed data. Admittedly the fuzzed data does appear to plateau earlier, with
a caveat however; the noise on the fuzzed data is such that there is quite possibly
a ``double approach'' to the plateau, due to contributions from more than one excited
state, as seen in the large statistics fuzzed data at $\beta=6.2$ in \fig{FigFuzzing},
so that with higher statistics the plateau may well not survive. 
It has also been found that the fuzzed-fuzzed combination tends to plateau after the
fuzzed-local, indicating opposing contributions from excited states, and meaning that
the plateau at early times is a misleading balance between contaminations of opposite sign
(at least until after the fuzzed-fuzzed correlator has reached a plateau), c.f. \fig{FigFuzzing}.

\section{Conclusions}

This smearing method provides, for the first time, a gauge invariant
technique for inserting arbitrary radial wavefunctions in a smeared operator.
So far results have only been presented using hydrogenic wavefunctions, though
in principle any form could be used. Excellent stability of fits to 
both the ground and radially excited states was demonstrated, and the 
level of statistical error was significantly smaller than traditional 
techiques on the same ensemble. Tuning the radius of the wavefunctions allowed
a clear plateau for the radially excited state to be isolated. An understanding
of the dependence of the fitted amplitudes on the radius of the smearing was
obtained, which can be used to 
determine more accurately the optimal radius.

The feasibility study did not include a sufficient number of quark masses to extrapolate to
the physical spectrum, however a rudimentary comparison of the $2S-1S$ splittings with
experiment is made in \tab{TabRadExpComp}, and can be seen to be plausible.
The main source of error, particularly in the light-light system,
is expected to be finite volume effects, 
since the Bohr radius required suggests the extent of the wavefunction
is significantly bigger than the lattice used. 

Since the technique allows a free choice for the wavefunction it would be interesting
to try to find a better basis than the hydrogenic wavefunctions. Harmonic oscillator
wavefunctions have been tried and found to be inferior to hydrogenic. Other
possible options are numerically solving the Cornell potential and inserting the 
solution as the smearing function, as has been performed
by the SESAM collaboration in the gauge fixed context \cite{SESAM_NRQCD}, 
or alternatively measuring the wavefunction
within a lattice calculation and re-inserting this. 
It would be interesting to compute the $R_{20}$ smeared correlators
on a large volume for light quarks since we expect finite volume effects on the
current volume.

\section{Acknowledgements}
The work of this paper was carried out
under the supervision of Richard Kenway, Ken Bowler 
and Brian Pendleton, whom I wish to thank for many useful
conversations. I also wish to thank 
Christine Davies for proof reading this work. 
The calculations were performed on the Cray T3D 
at the EPCC using UKQCD computer time.
I wish to acknowledge the support of EPSRC grant  GR/K41663 and PPARC
grant GR/K55745. 
The author was funded by the Carnegie Trust for the Universities of Scotland
while this work was carried out at the University of Edinburgh,
and is grateful for PPARC grant PP/CBA/62, 
and to the University of Glasgow for support while writing this paper.

\fltfig{Heavy heavy pseudoscalar fits for $r_0 = 2.4, 2.5, 2.6$. 
The plots contain the effective masses in lattice units 
for correlators with $R_{20}$ (circles) and $R_{10}$ (fancy squares) 
source smearings with local sinks. The fits are simultaneous double
exponential over the timeslice range 4-20.
A clear transition between a pole in the $R_{20}$ effective mass and
a continuous effective mass is seen as $r_0$ is tuned across its
optimal value.
}{
\epsfaxhaxhax
{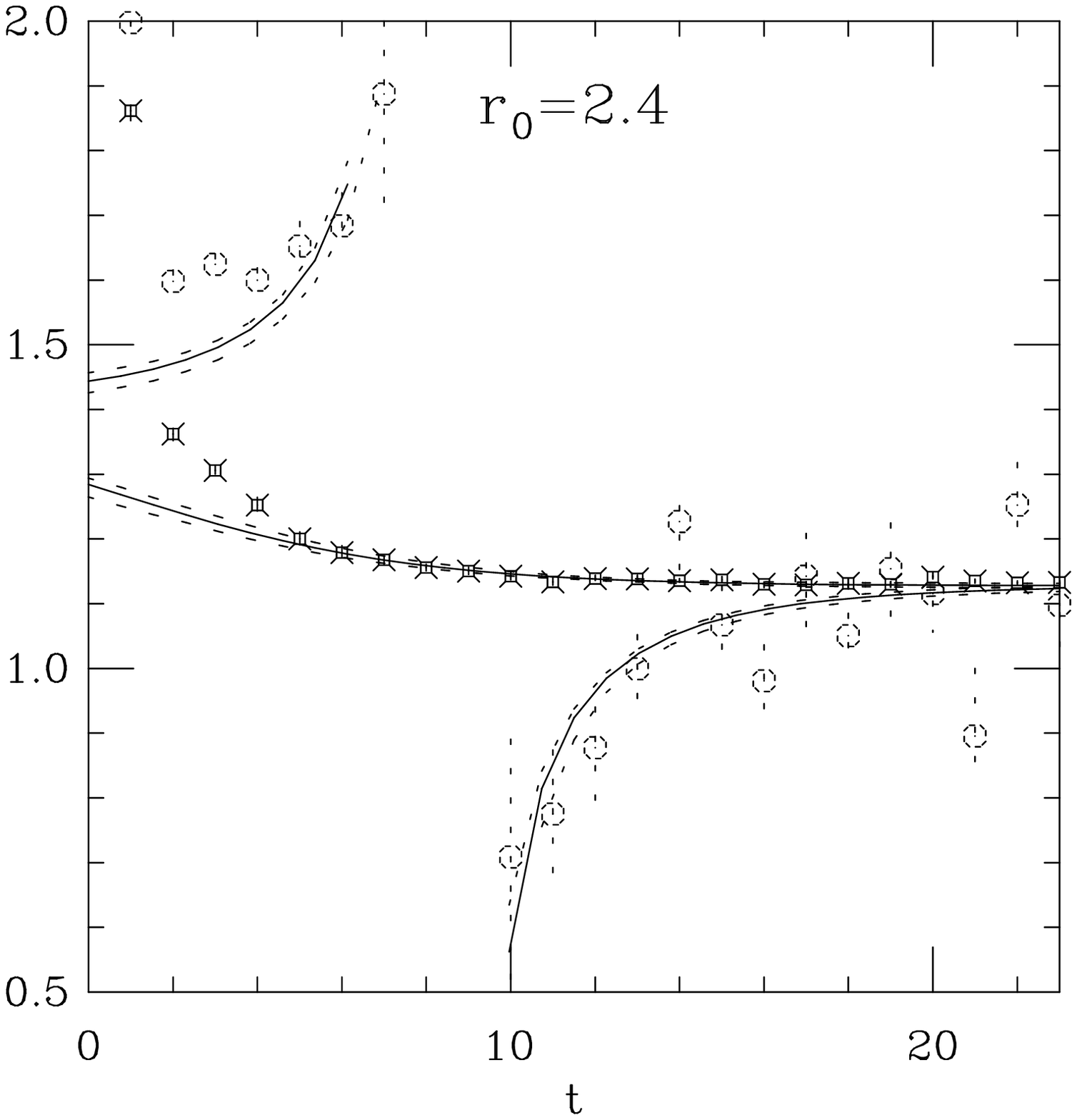}
{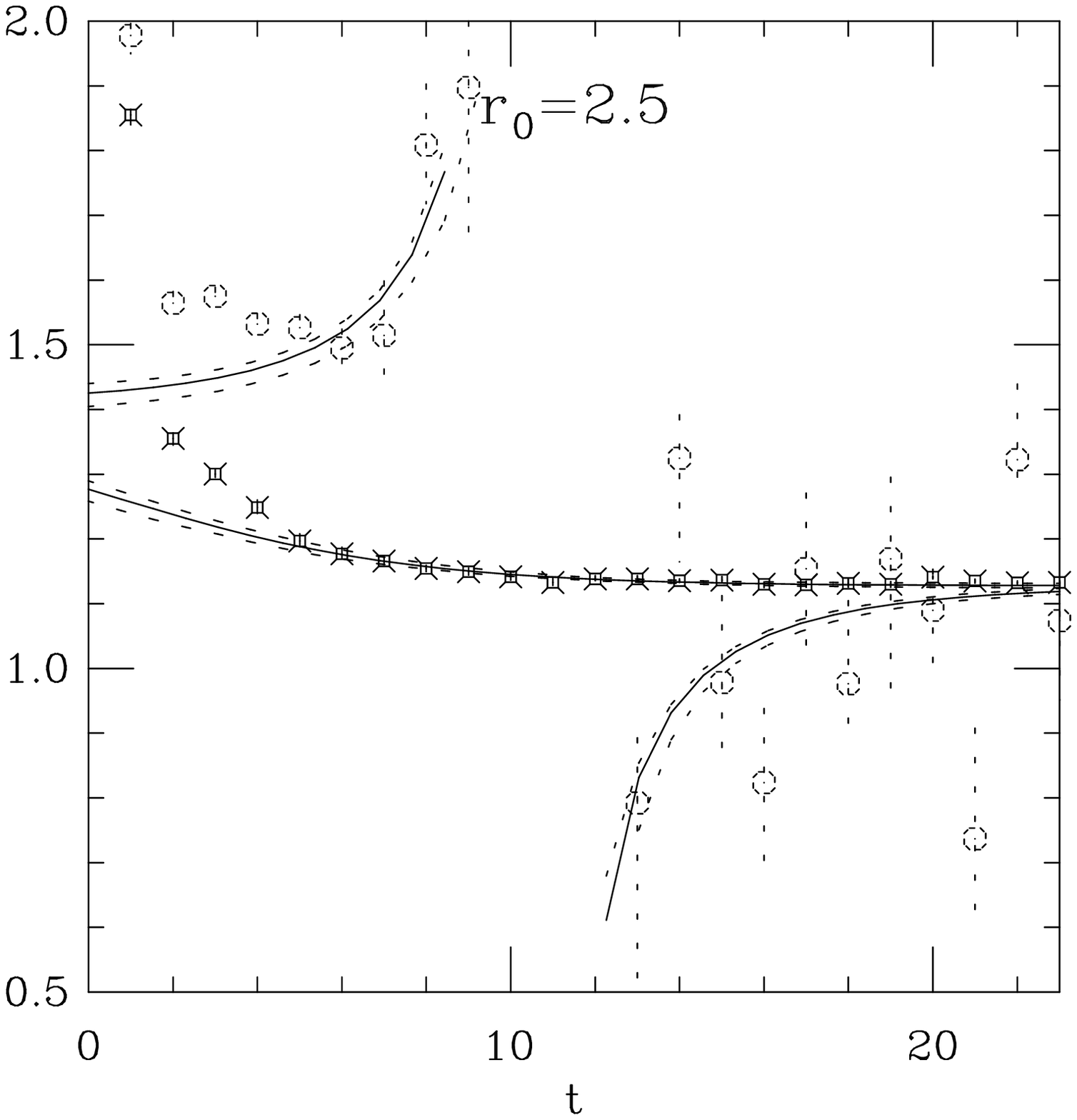}
{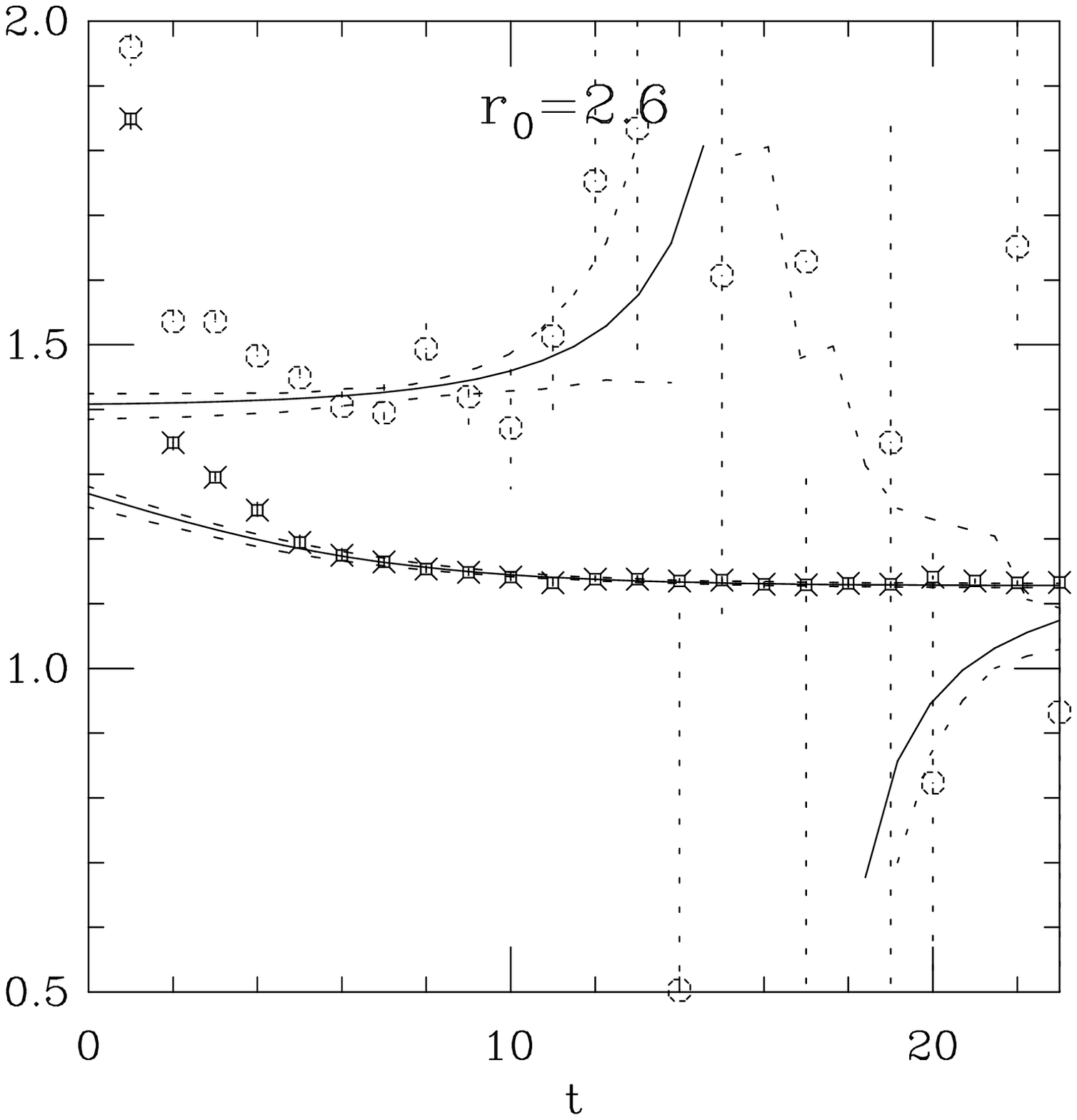}
}{FIGBOYLE1}

\fltfig{Heavy heavy pseudoscalar fits $r_0 = 2.75, 3.0, 5.5$
The plots contain the effective masses in lattice units 
for correlators with $R_{20}$ (circles) and $R_{10}$ (fancy squares) 
source smearings with local sinks. The fits are simultaneous double
exponential over the timeslice range 4-20.
}{
\epsfaxhaxhax
{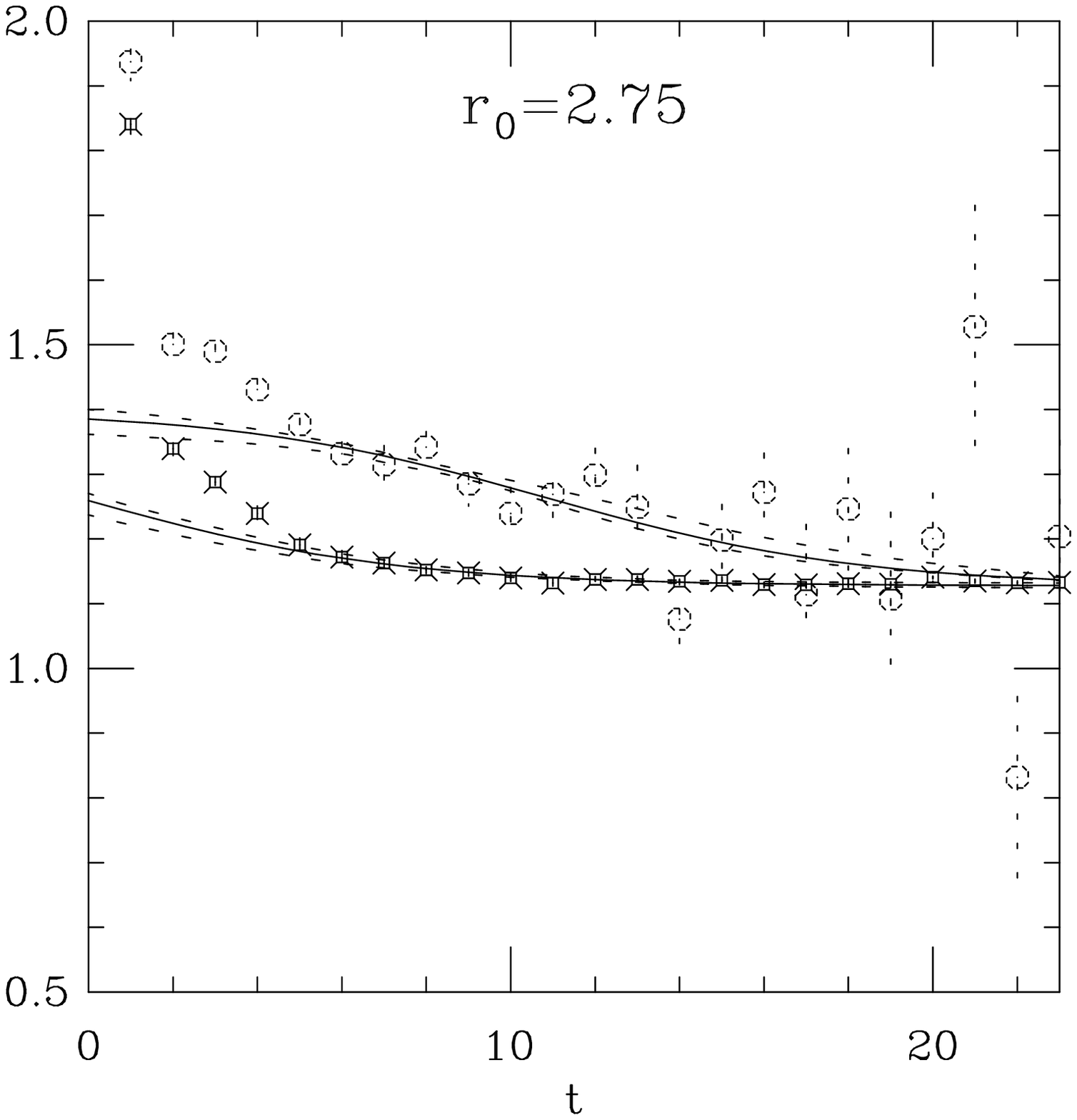}
{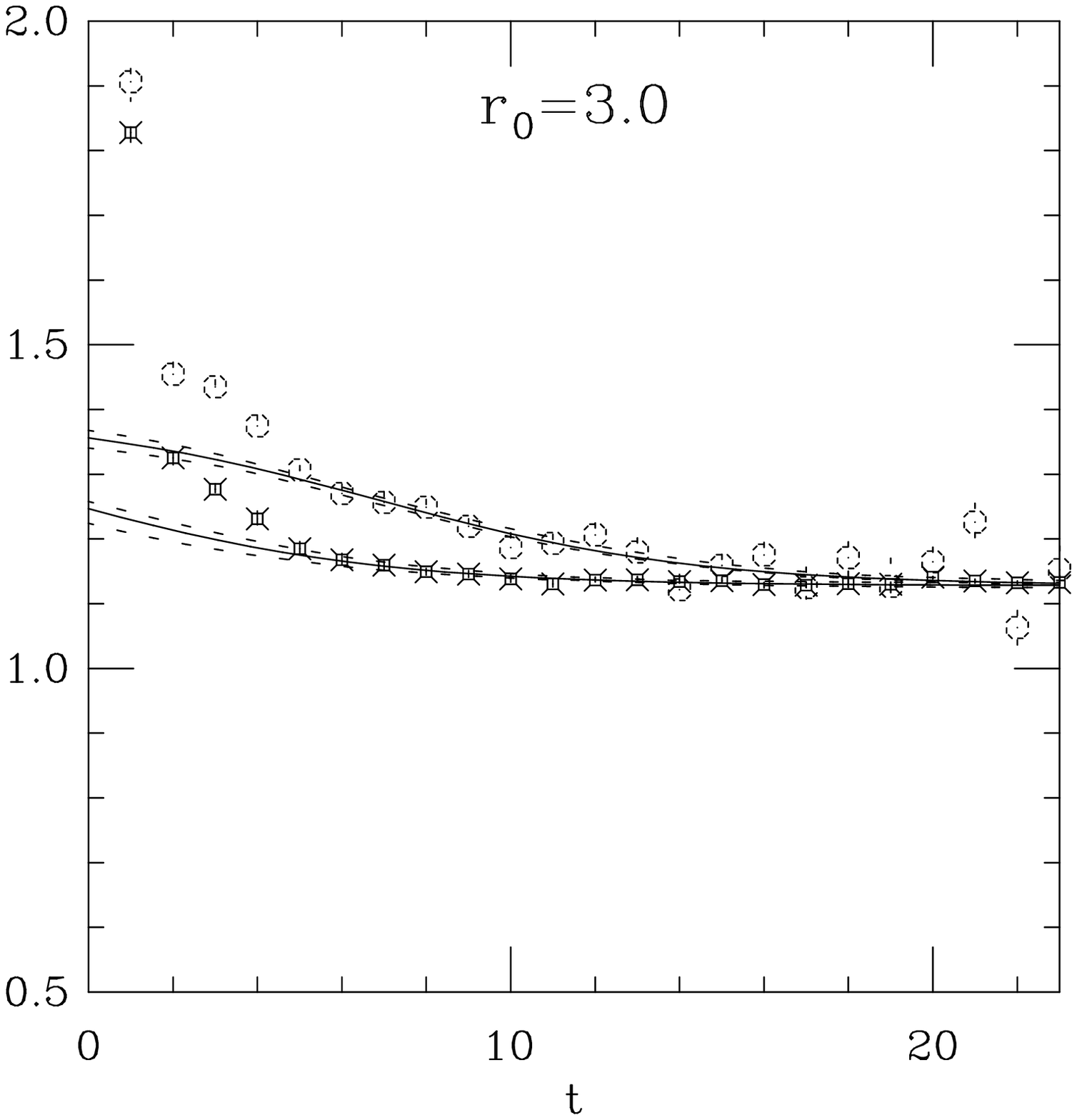}
{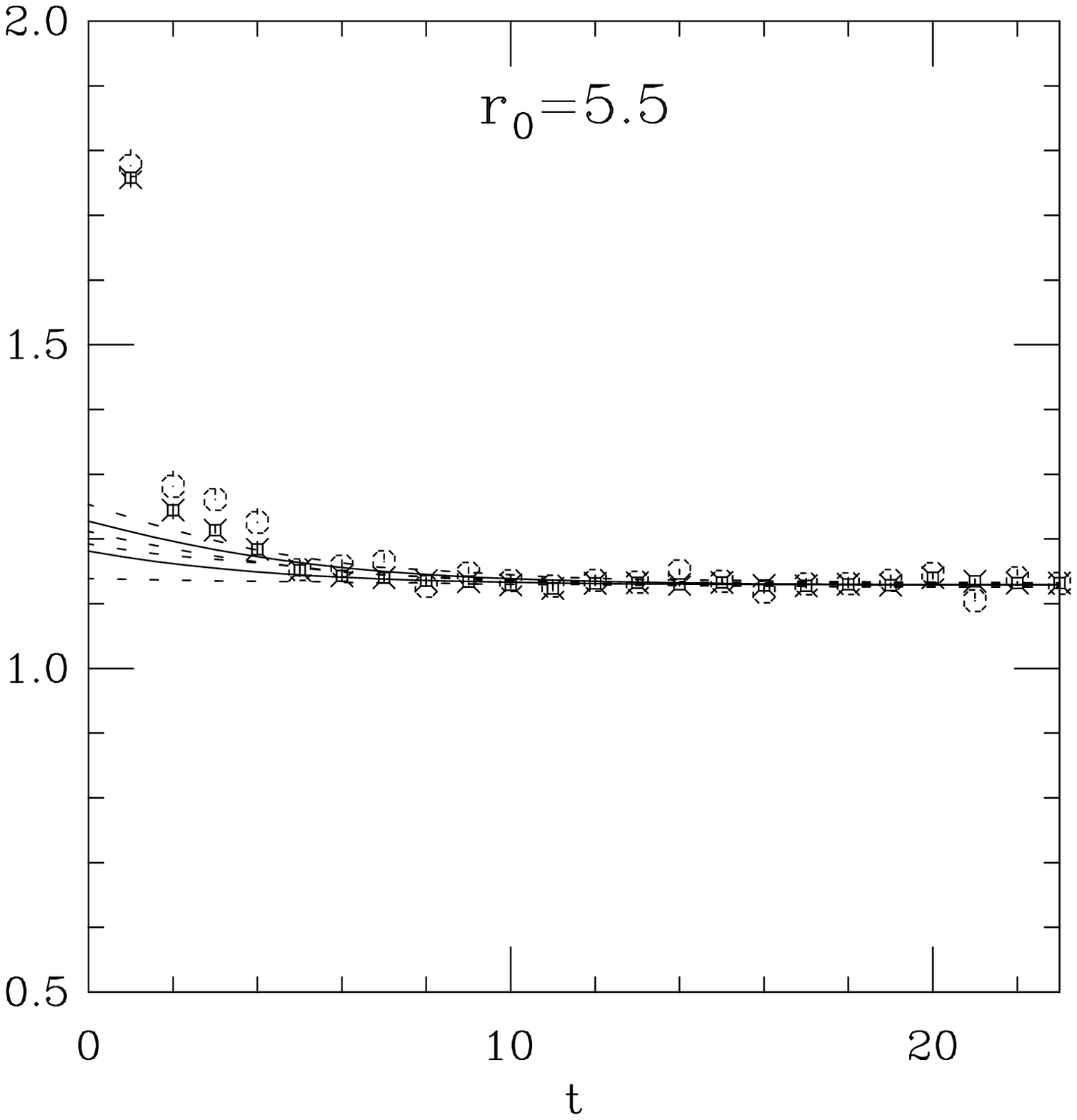}
}{FIGBOYLE2}

\fltfig{Heavy light pseudoscalar fits, $r_0 = 2.6, 2.8, 2.9$.
The plots contain the effective masses in lattice units 
for correlators with $R_{20}$ (circles) and $R_{10}$ (fancy squares) 
source smearings with local sinks. The fits are simultaneous double
exponential over the timeslice range 4-20.
}{
\epsfaxhaxhax
{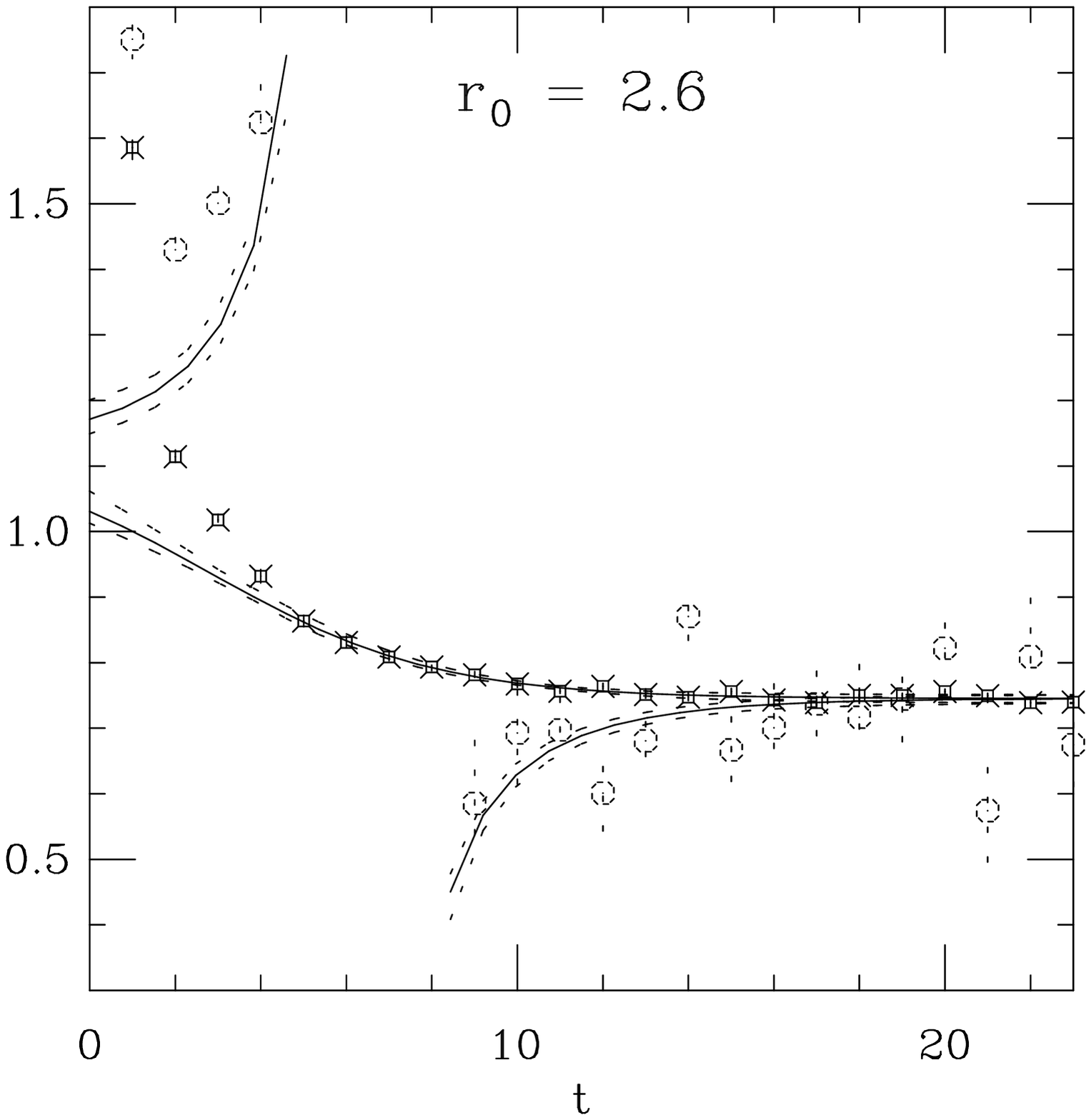}
{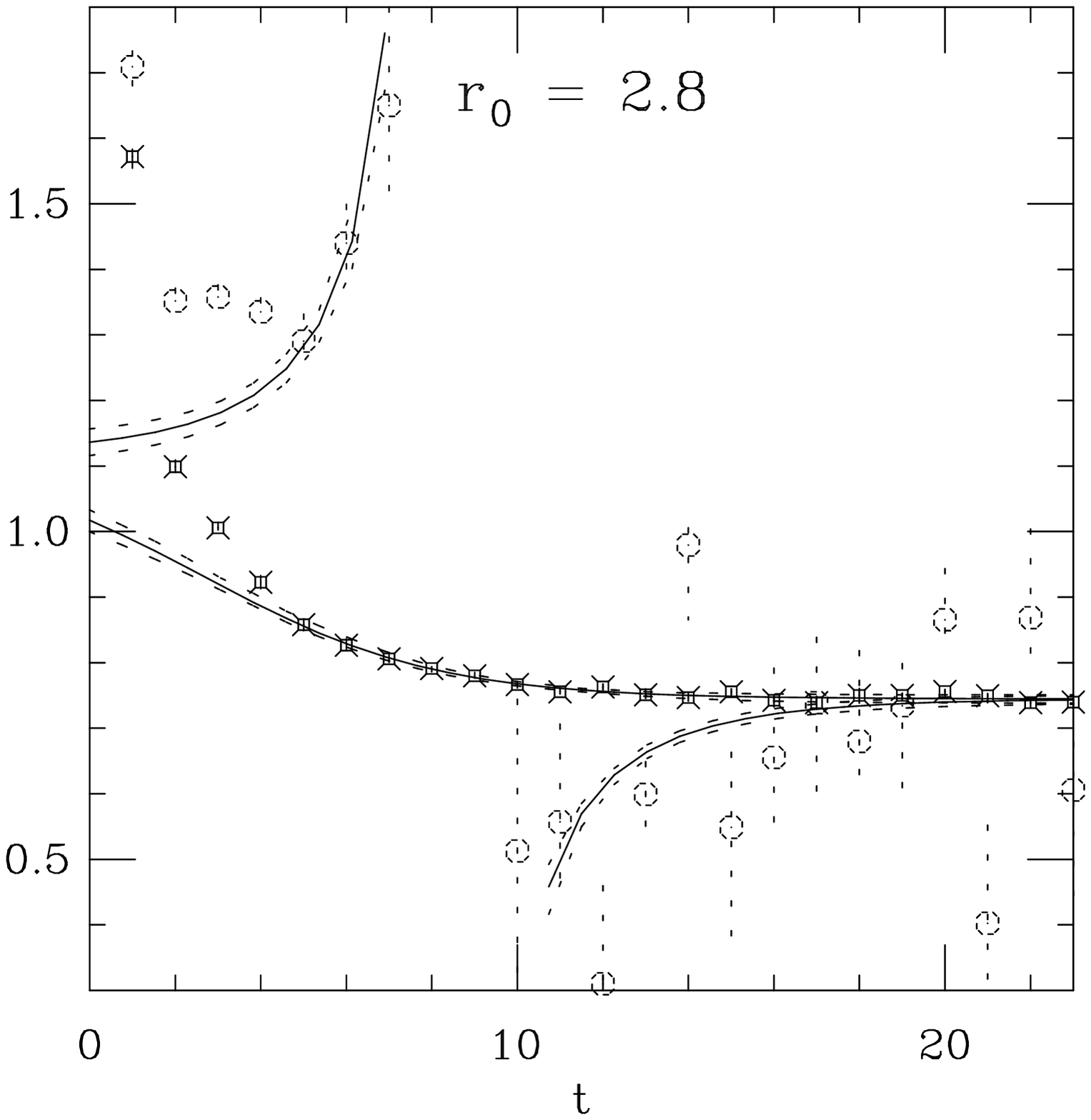}
{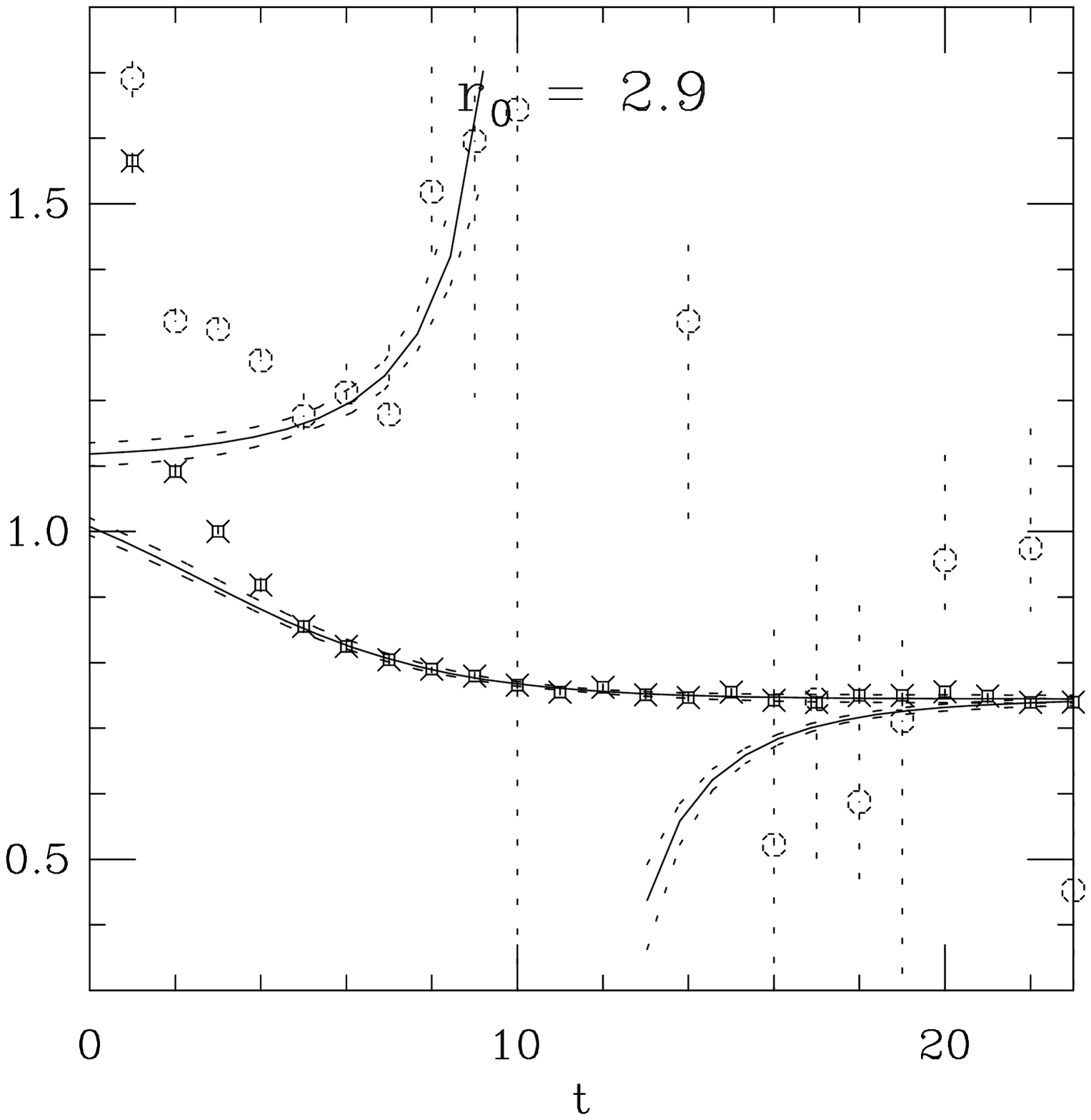}
}{FIGBOYLE5}

\fltfig{Heavy light pseudoscalar fits $r_0 = 3.0, 3.1, 4.0$.
The plots contain the effective masses in lattice units 
for correlators with $R_{20}$ (circles) and $R_{10}$ (fancy squares) 
source smearings with local sinks. The fits are simultaneous double
exponential over the timeslice range 4-20.
}{
\epsfaxhaxhax
{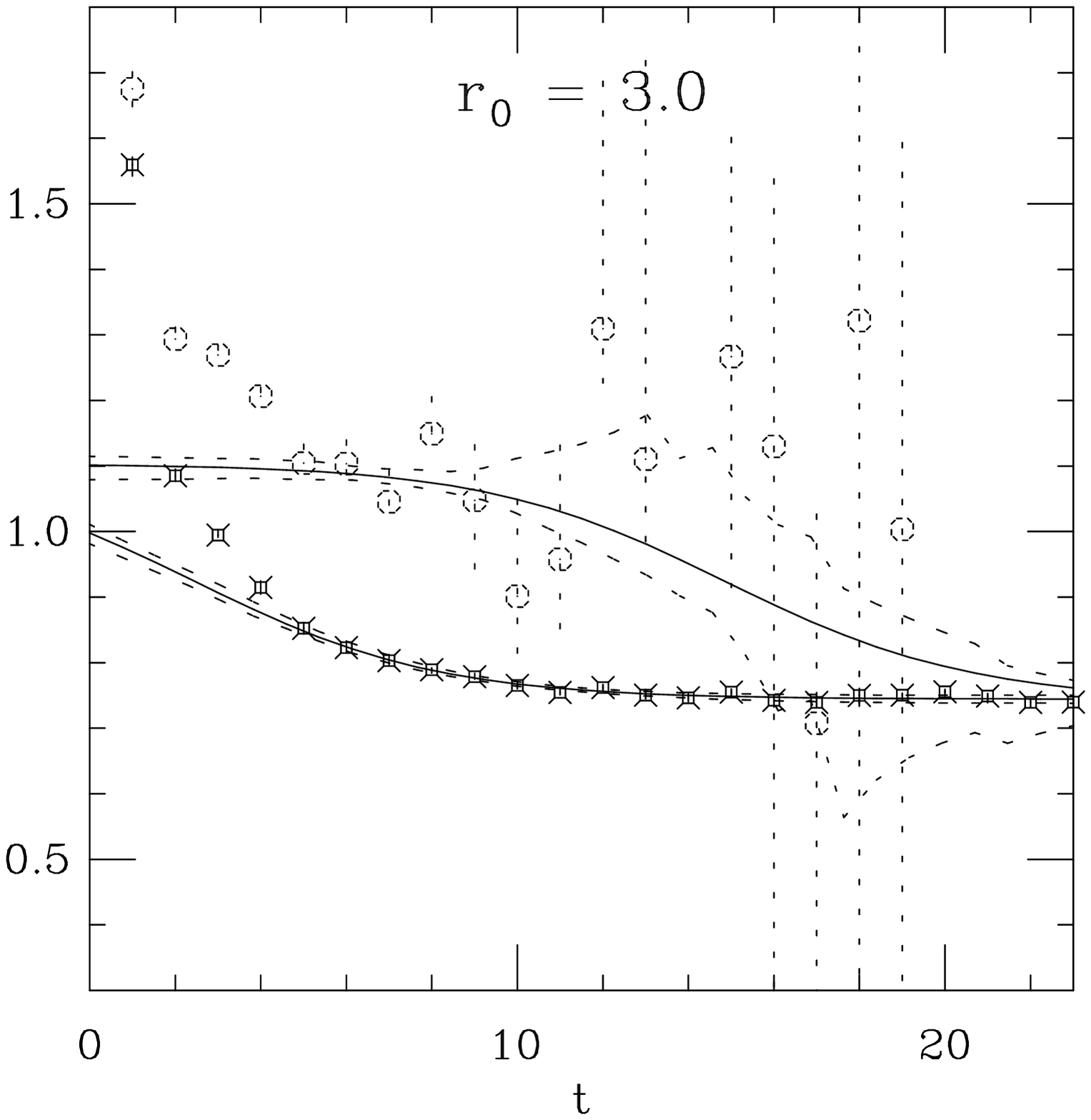}
{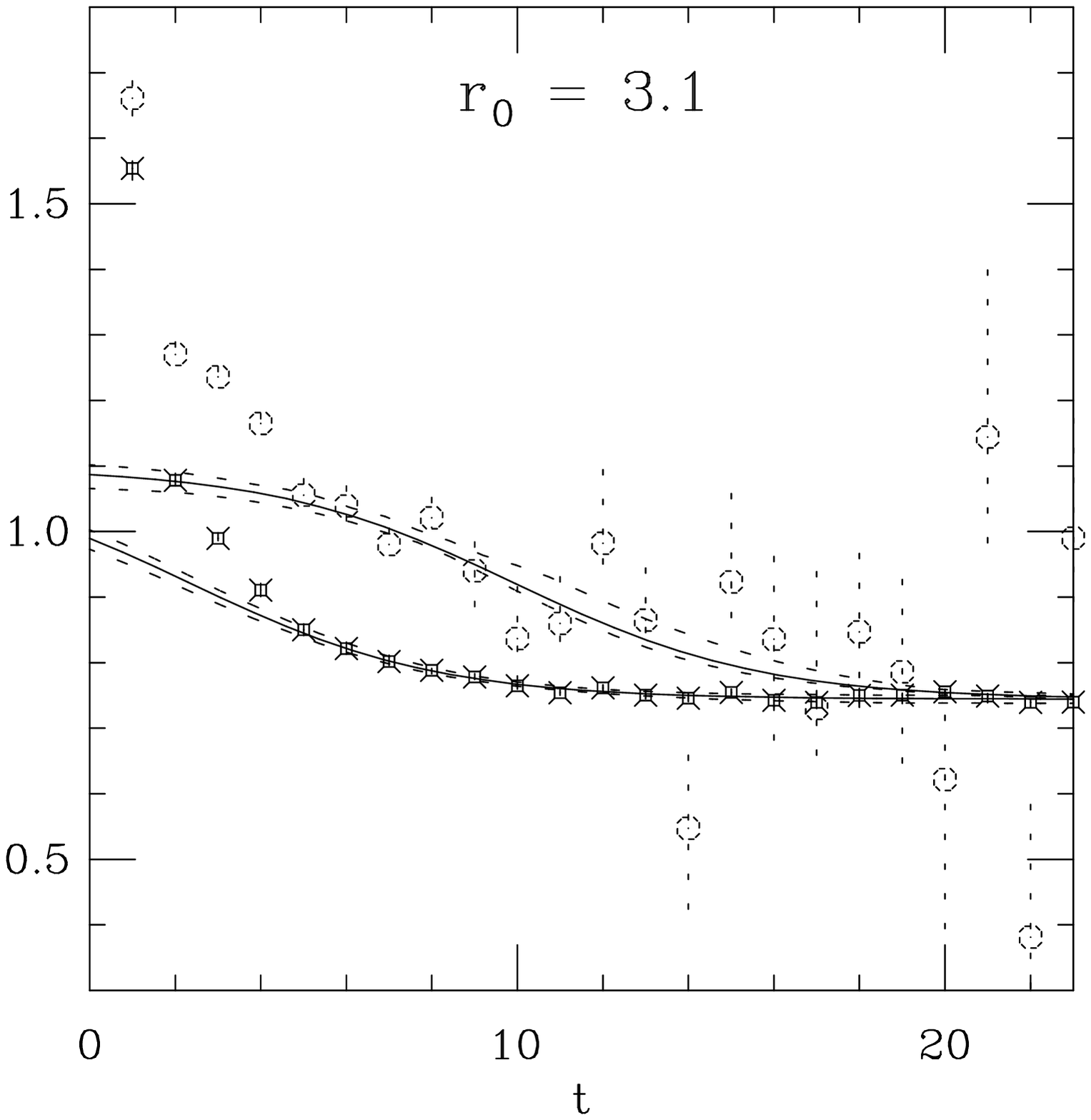}
{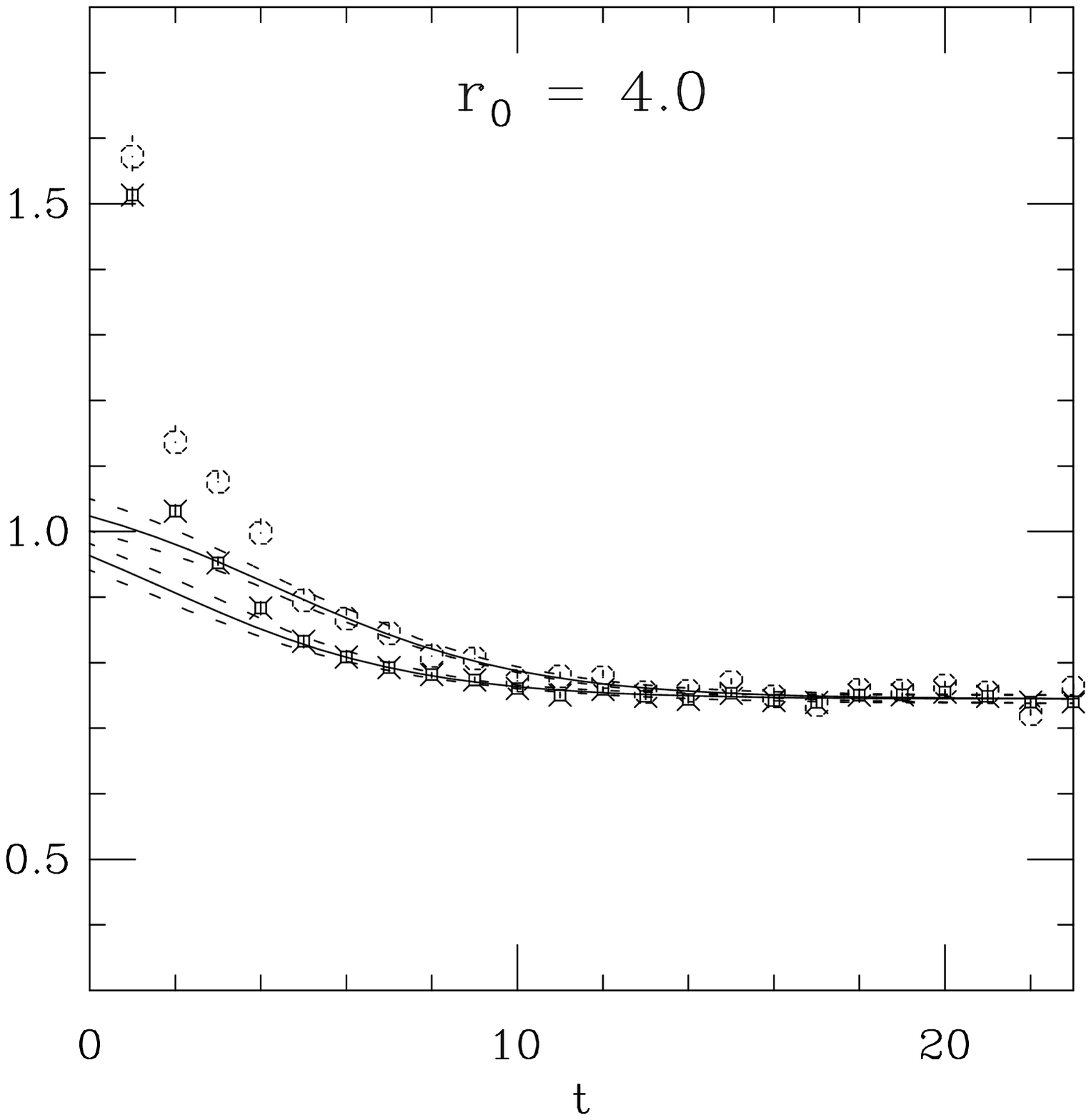}
}{FIGBOYLE6}

\fltfig{Light light pseudoscalar fits, $r_0 = 2.8, 2.9, 3.0$
The plots contain the effective masses in lattice units 
for correlators with $R_{20}$ (circles) and $R_{10}$ (fancy squares) 
source smearings with local sinks. The fits are simultaneous double
exponential over the timeslice range 4-18.
}{
\epsfaxhaxhax
{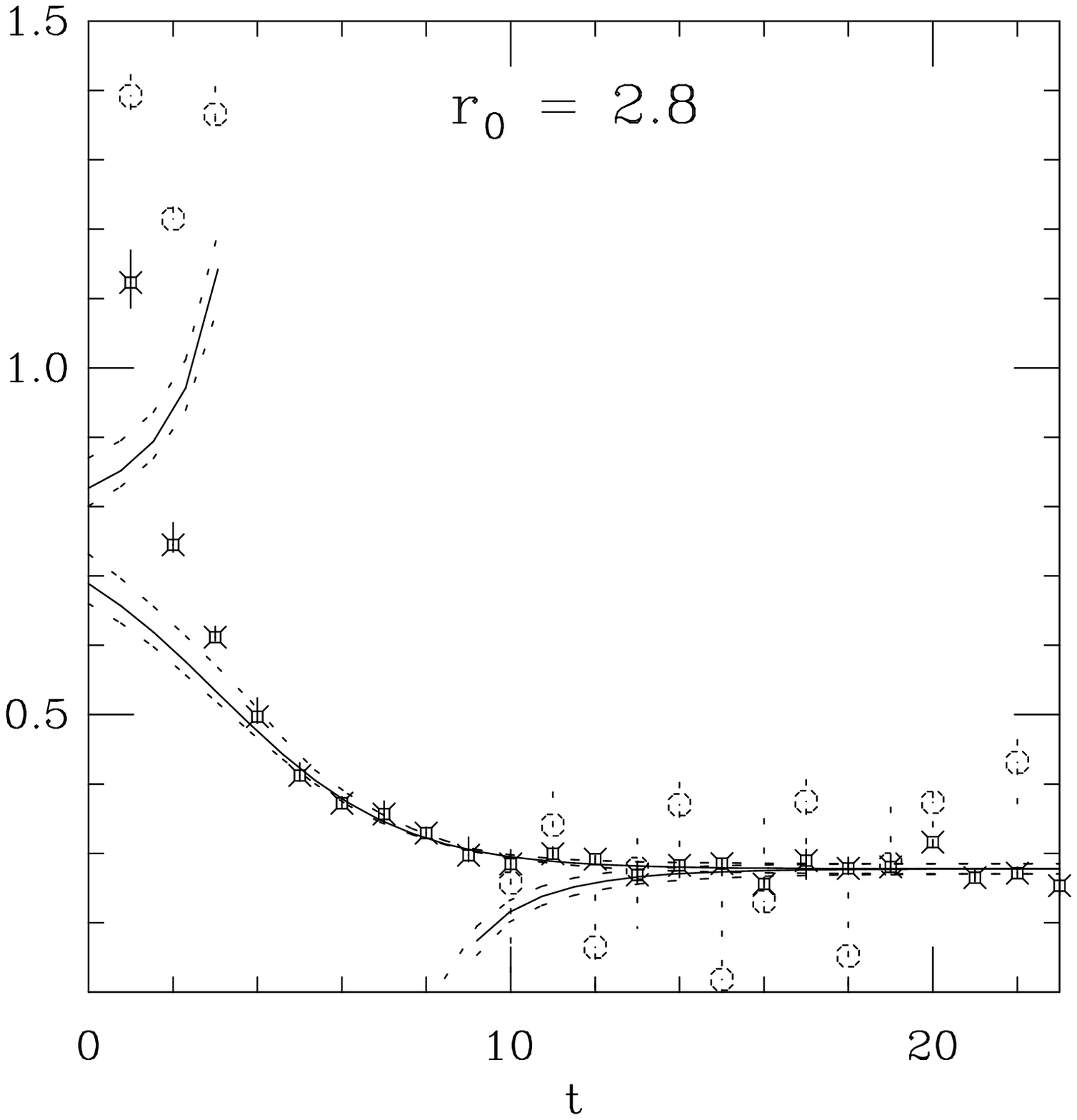}
{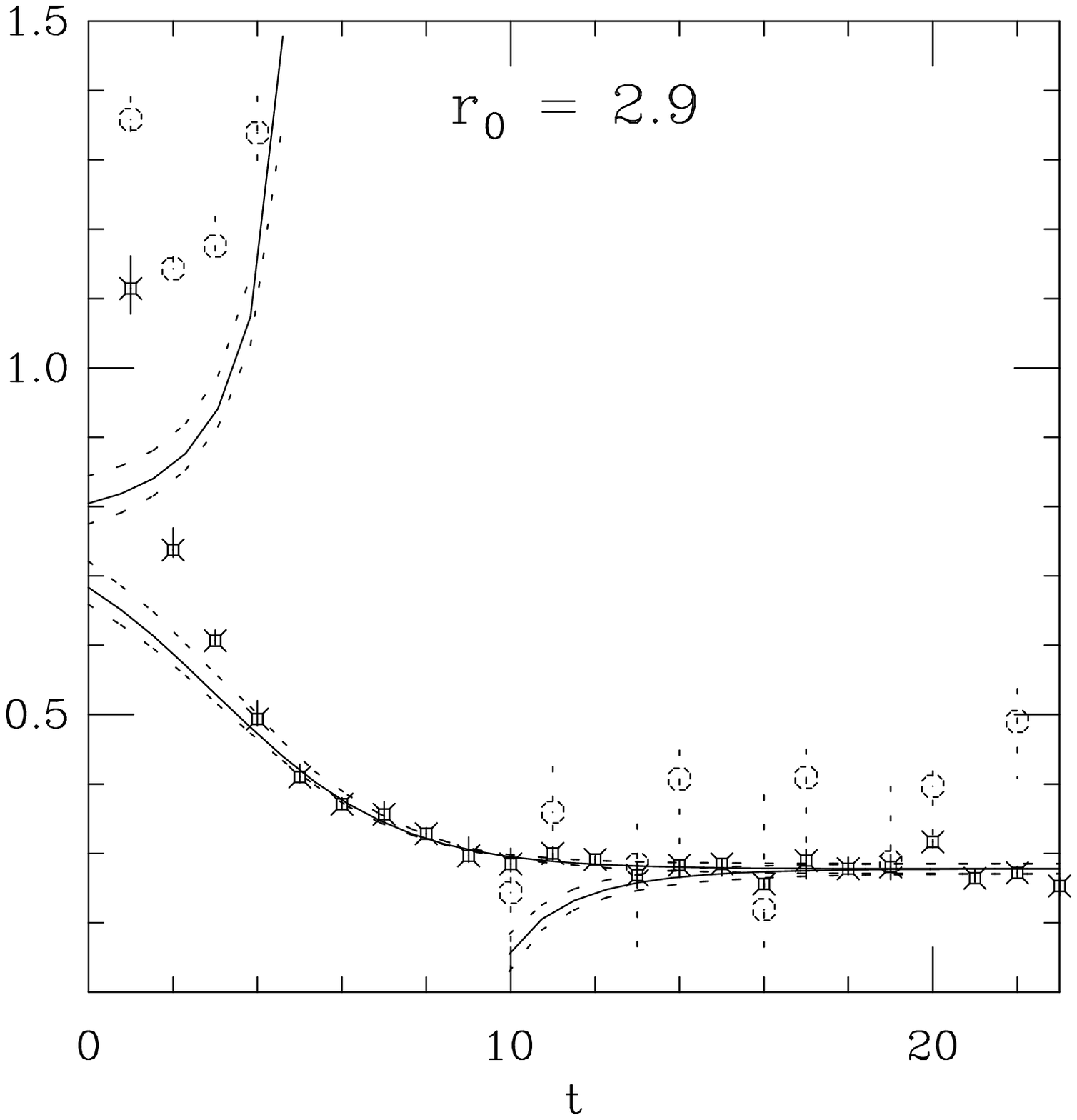}
{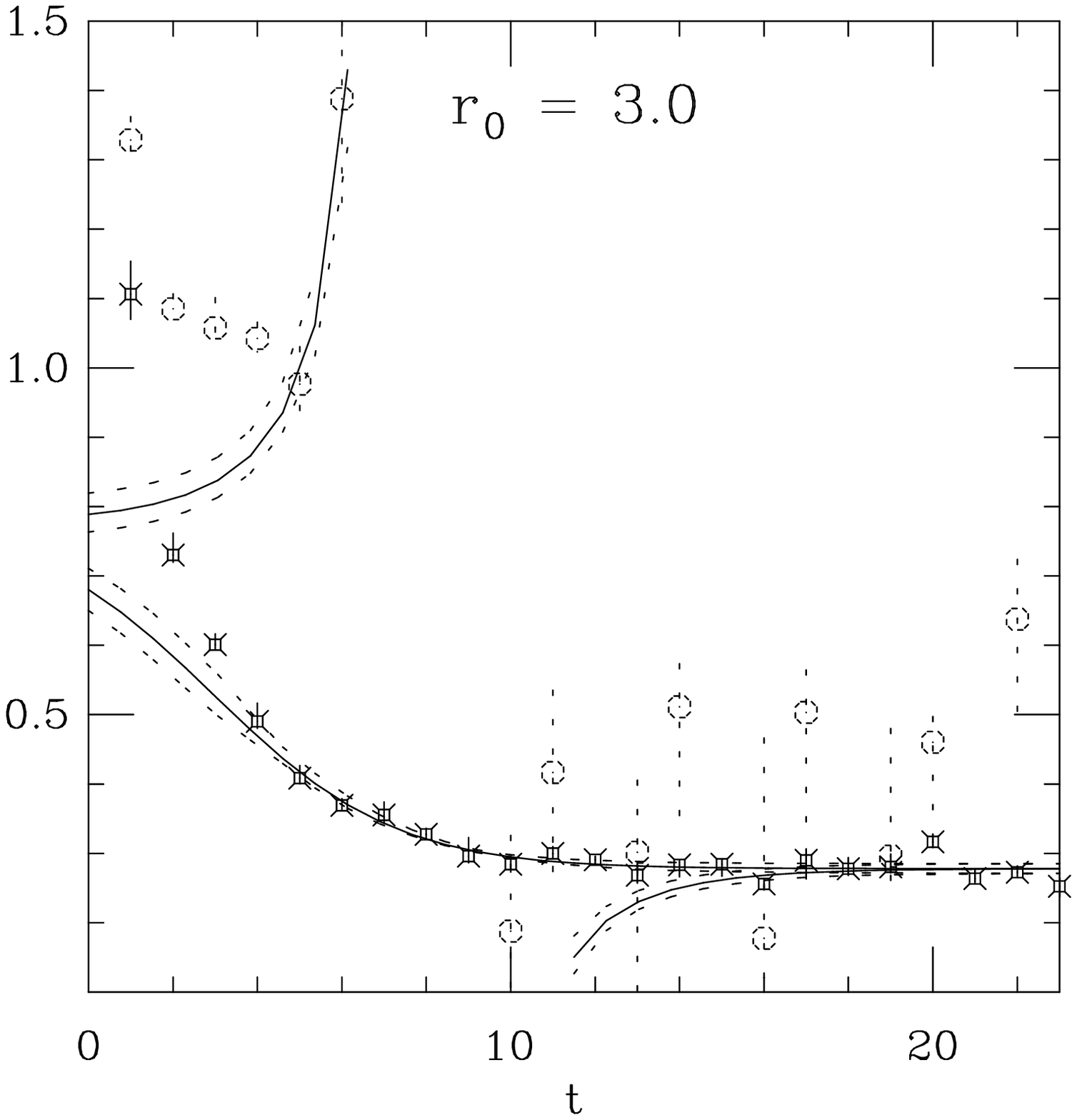}
}{FIGBOYLE9}

\fltfig{Light light pseudoscalar fits $r_0 = 3.1, 3.25, 4.0$.
The plots contain the effective masses in lattice units 
for correlators with $R_{20}$ (circles) and $R_{10}$ (fancy squares) 
source smearings with local sinks. The fits are simultaneous double
exponential over the timeslice range 4-18.
}{
\epsfaxhaxhax
{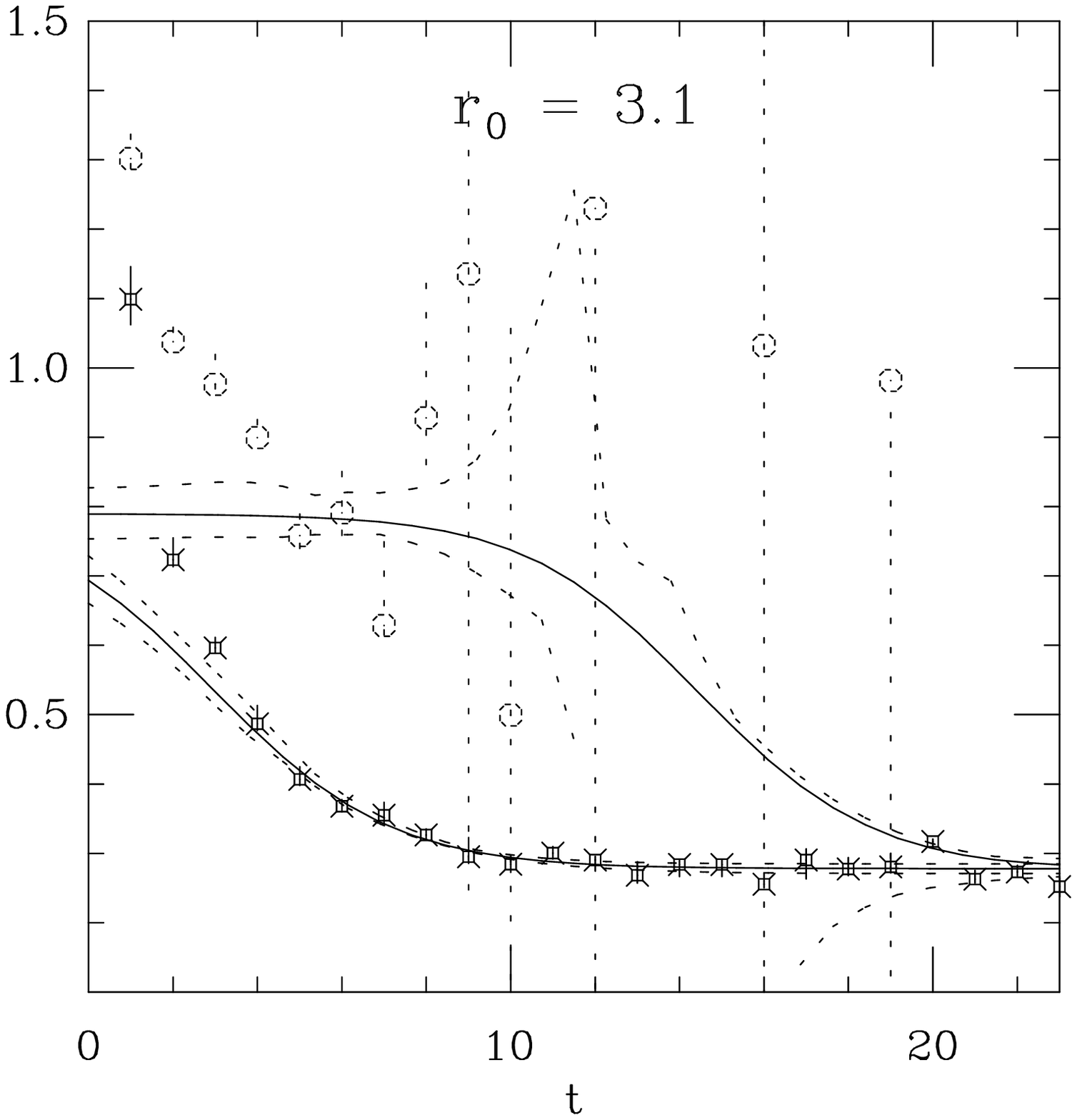}
{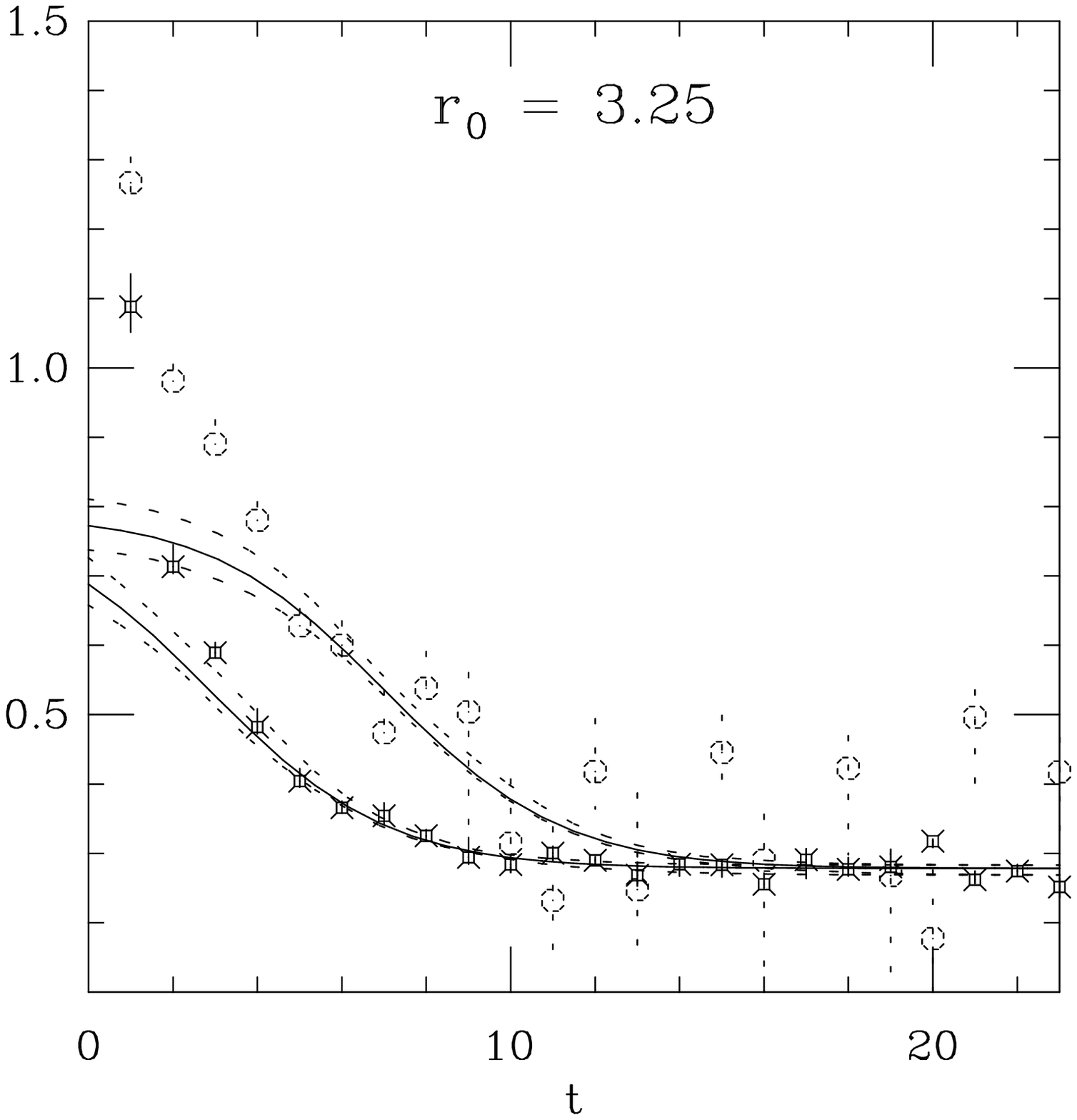}
{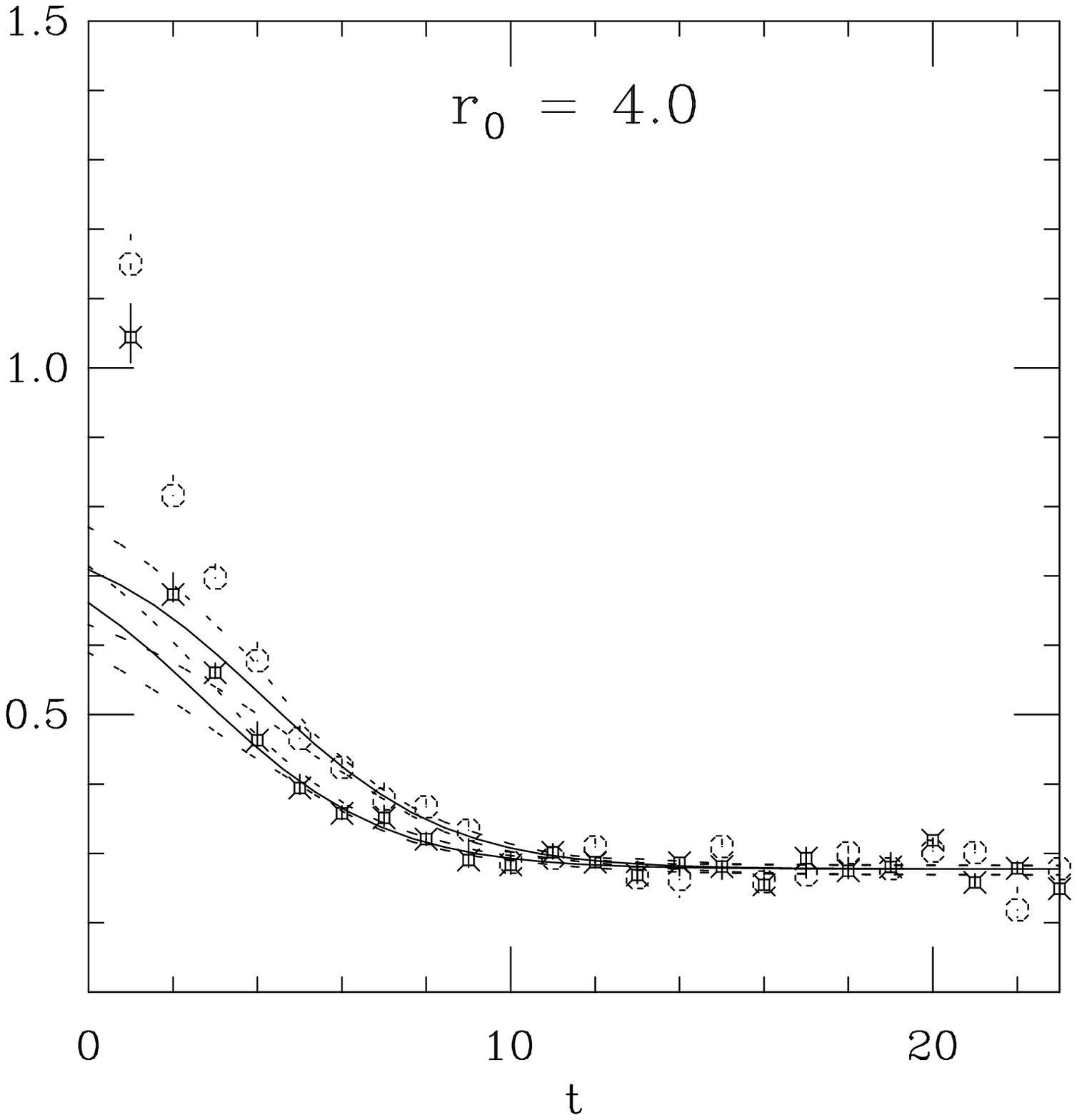}
}{FIGBOYLE11}

\fltfig{Fitted amplitudes for quarkonia correlation functions using the
timeslice range 4-20. The contaming amplitude in the $R_{20}$ source smeared
correlator clearly shows a zero at $r_0 \simeq 2.6$}
{\pspicture{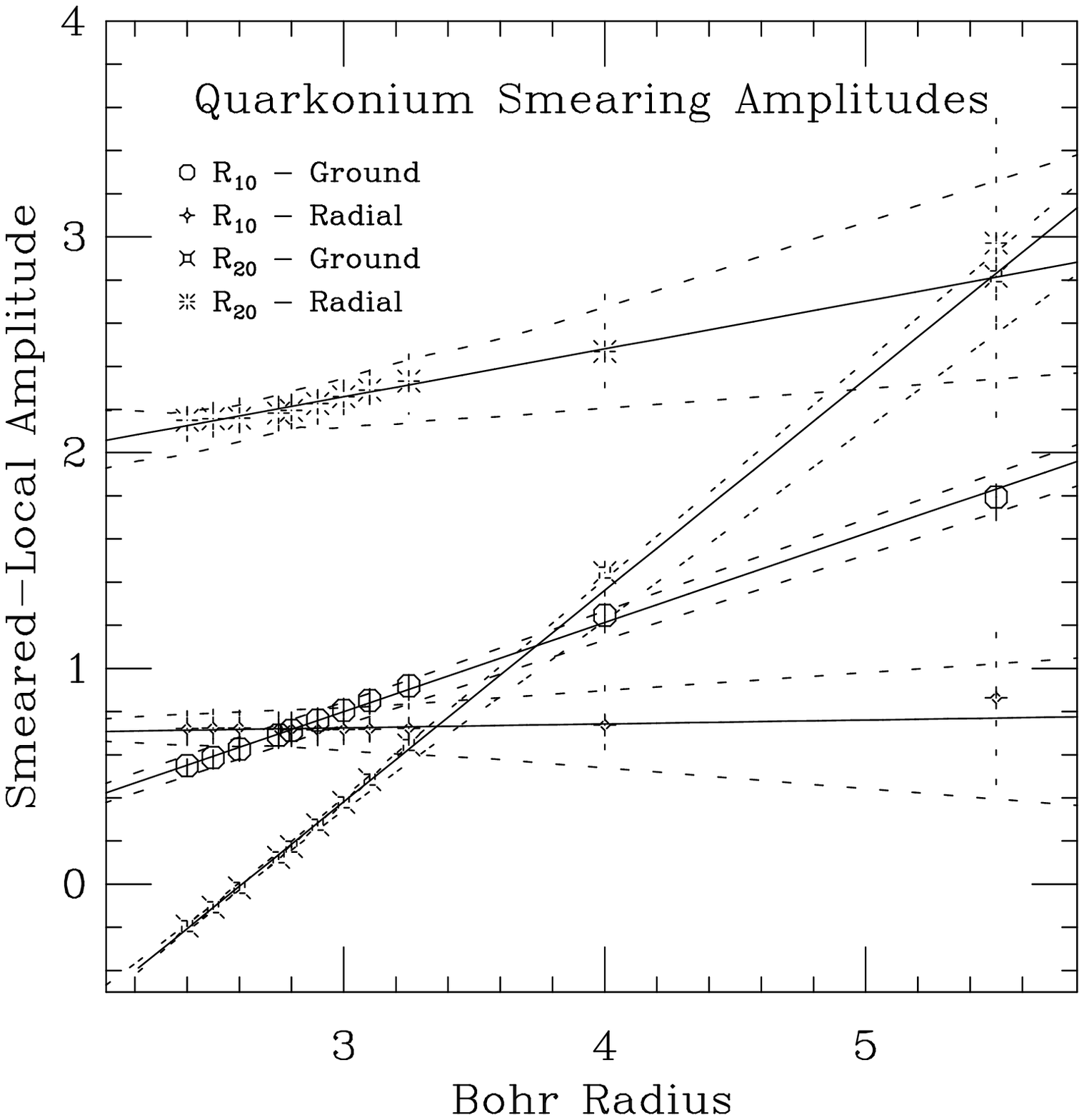}}
{FigHHAmps}
\fltfig{Fitted amplitudes for heavy-light correlation functions
using the timeslice range 4-20. 
The contaming amplitude in the $R_{20}$ source smeared
correlator clearly shows a zero at $r_0 \simeq 3.0$}
{\pspicture{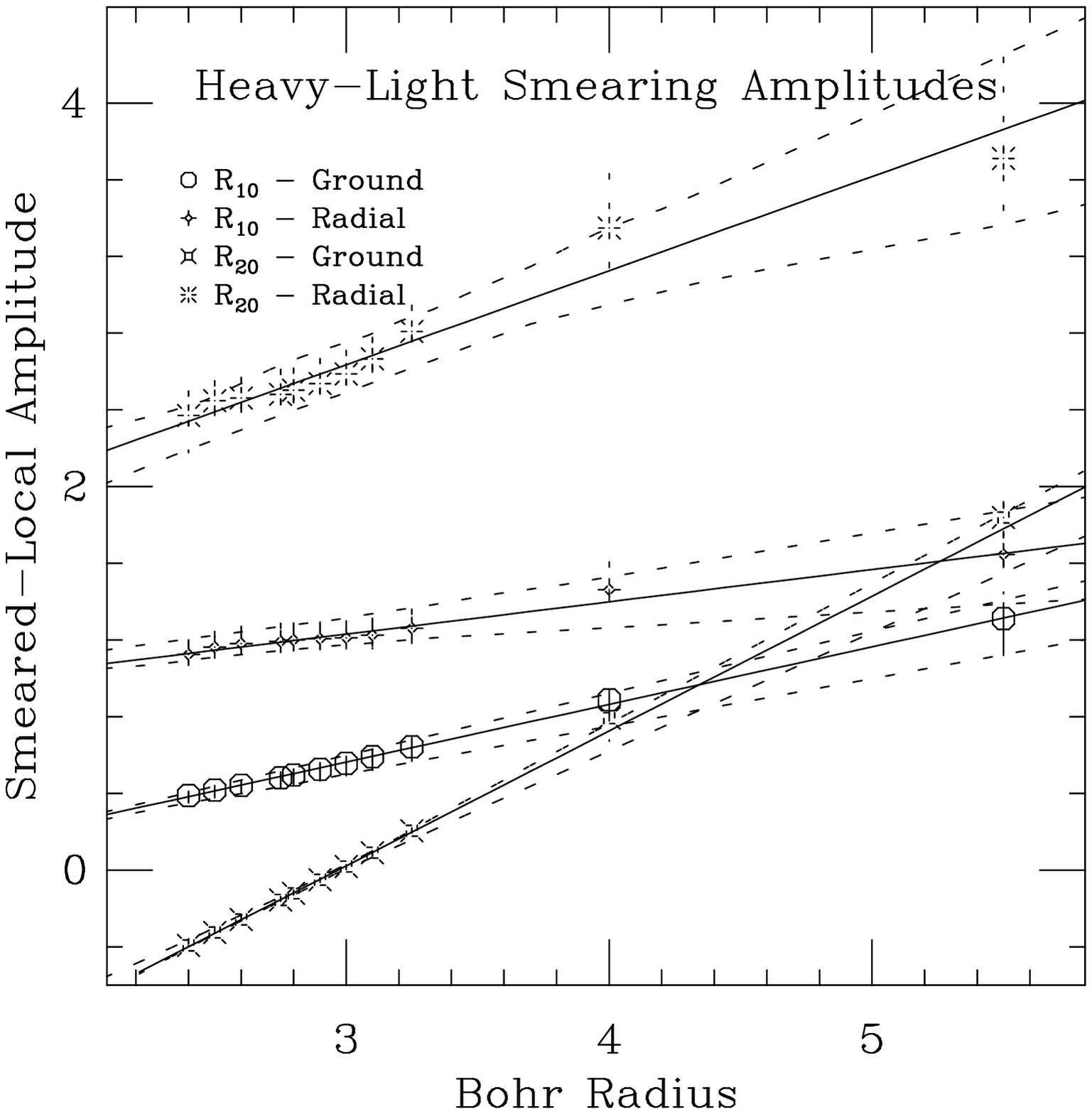}}
{FigHLAmps}
\fltfig{Fitted amplitudes for light-light correlation functions using
the timeslice range 4-20. 
The contaming amplitude in the $R_{20}$ source smeared
correlator clearly shows a zero at $r_0 \simeq 3.1$}
{\pspicture{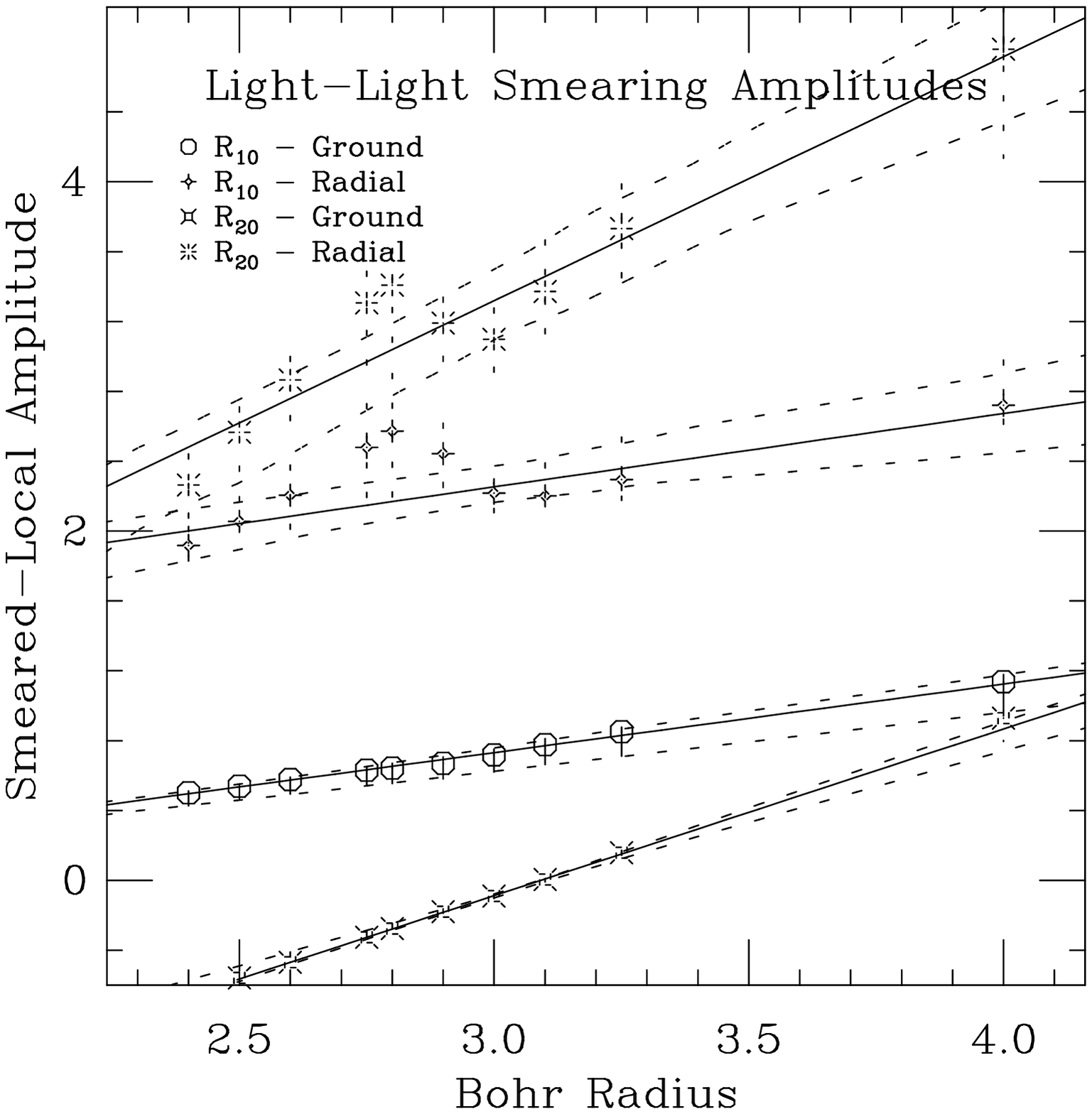}}
{FigLLAmps}

\fltfig{$R_{10}$ - $R_{10}$ source and sink
smeared quarkonium correlator (fancy squares), used in a double
exponential fit with the $R_{20}$ source, local sink
correlator (circles) with Bohr radius $r_0 = 2.4,3.0$}
{
\epsfaxhax{ 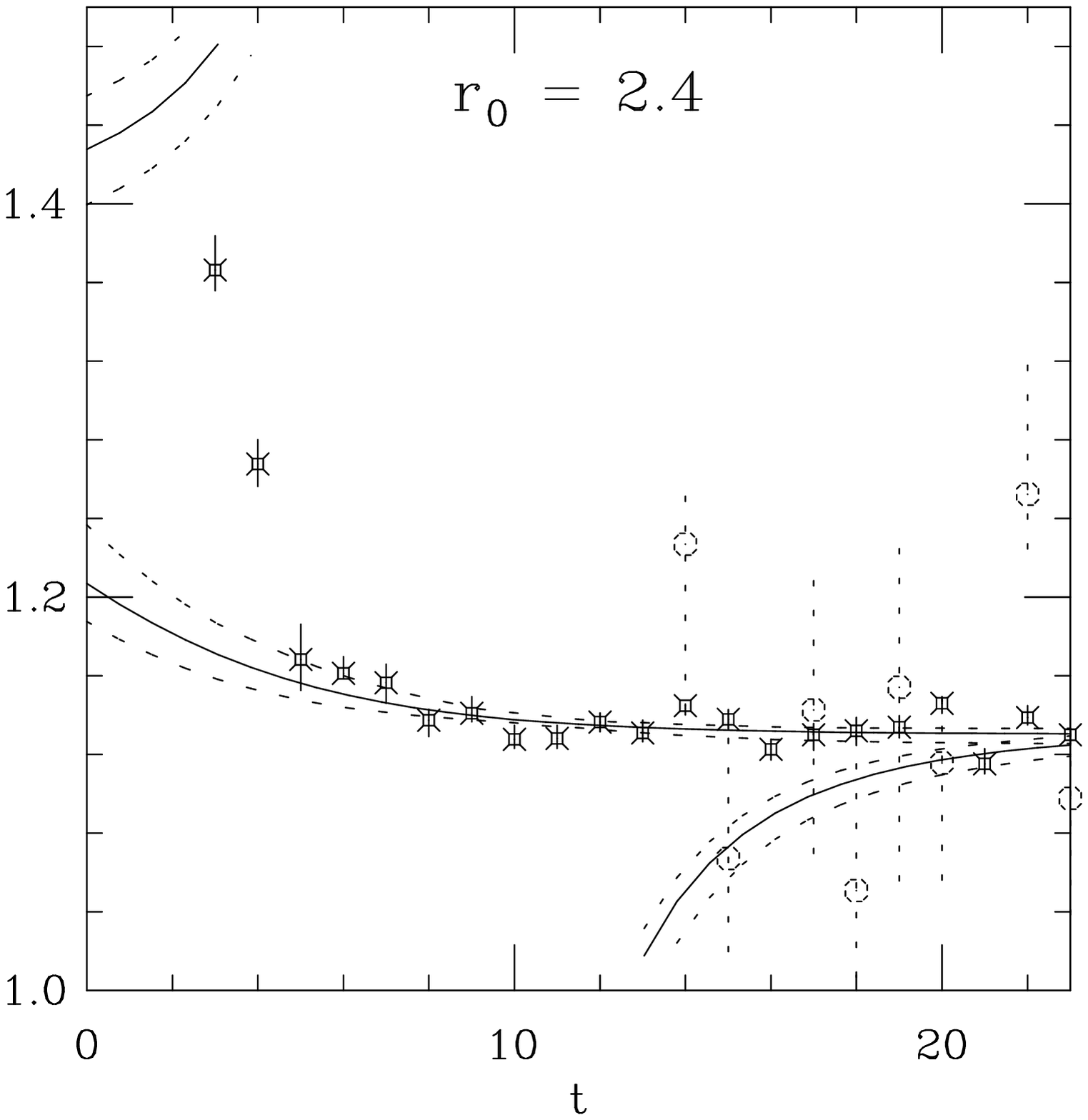}
{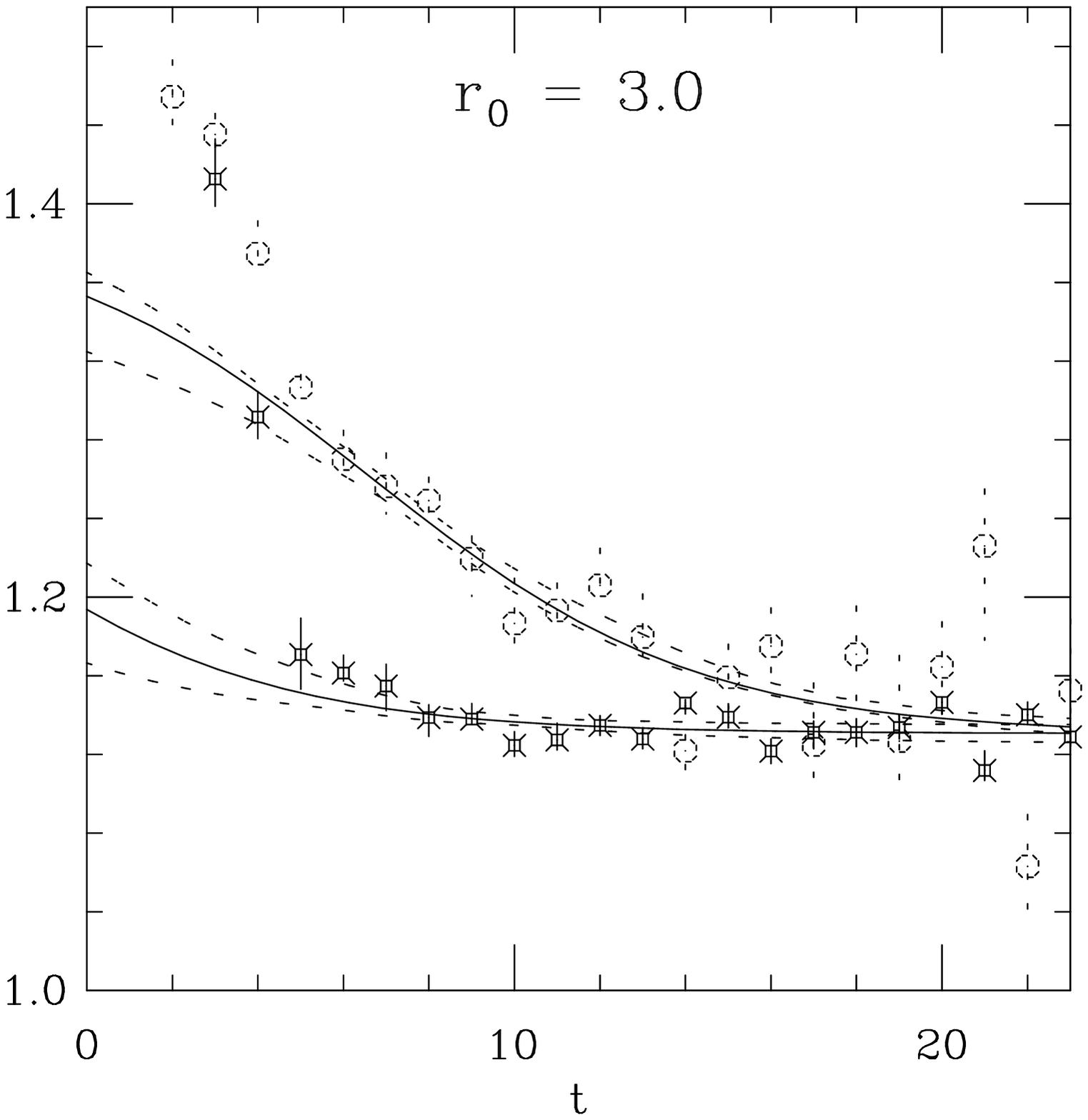}
}
{FigSScontamination}

\fltfig{Contaminating amplitude obtained from the 
$R_{10}$ - $R_{10}$ source and sink smeared quarkonium correlator.
This smearing radius minimising this amplitude lies
in the region of 2.1(4). 
}{
\pspicture{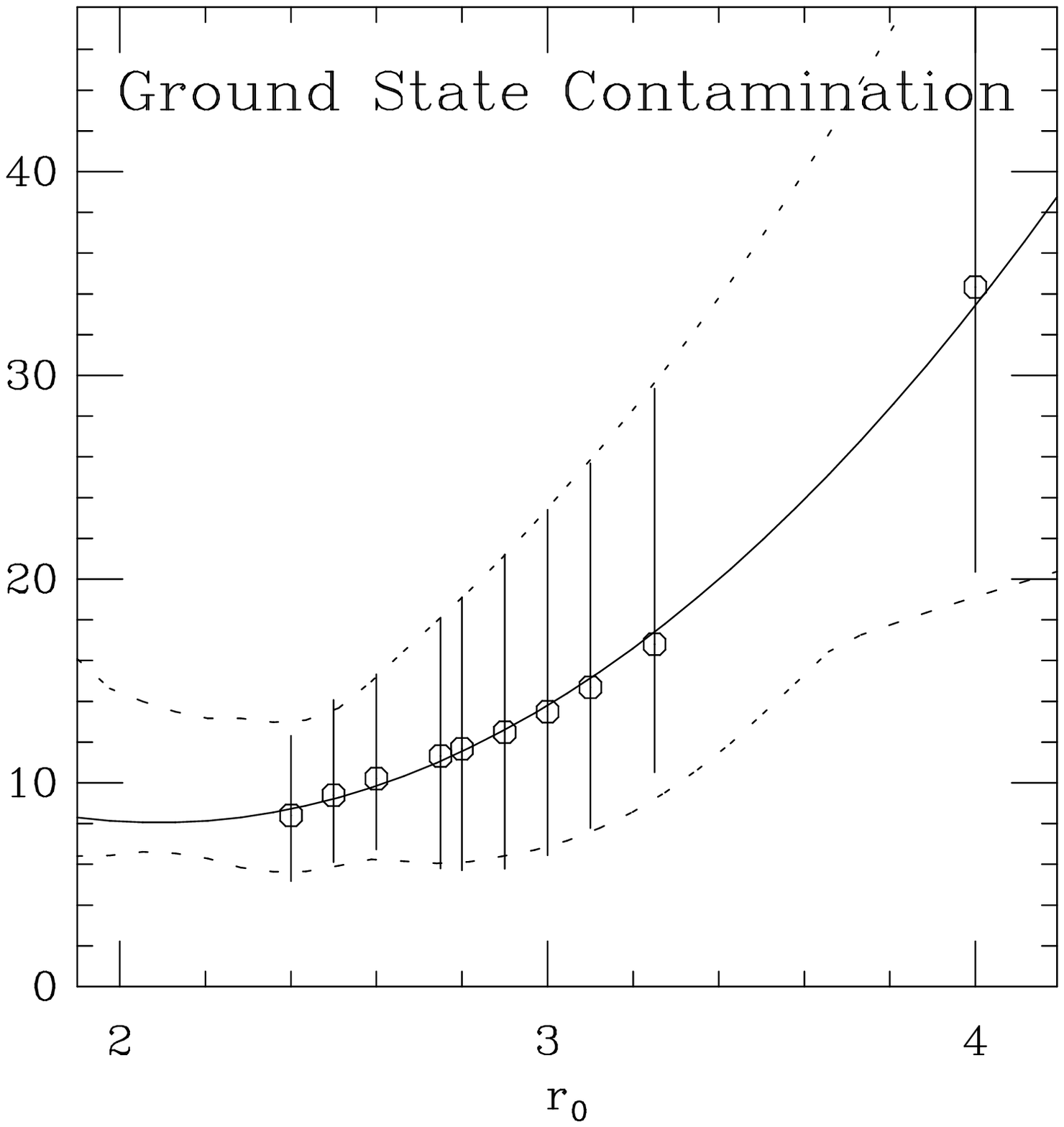}
}
{FigSSamp}

\fltfig{Fit to the doubly $R_{10}$ source smeared with local sink
correlator (fancy squares) and the $R_{20}$ source smeared with local
sink correlator (circles) for the heavy-heavy meson. The quality
of the plateau suggests that the use of double smearing is
worthwhile for additional information on the groundstate.
}
{\pspicture{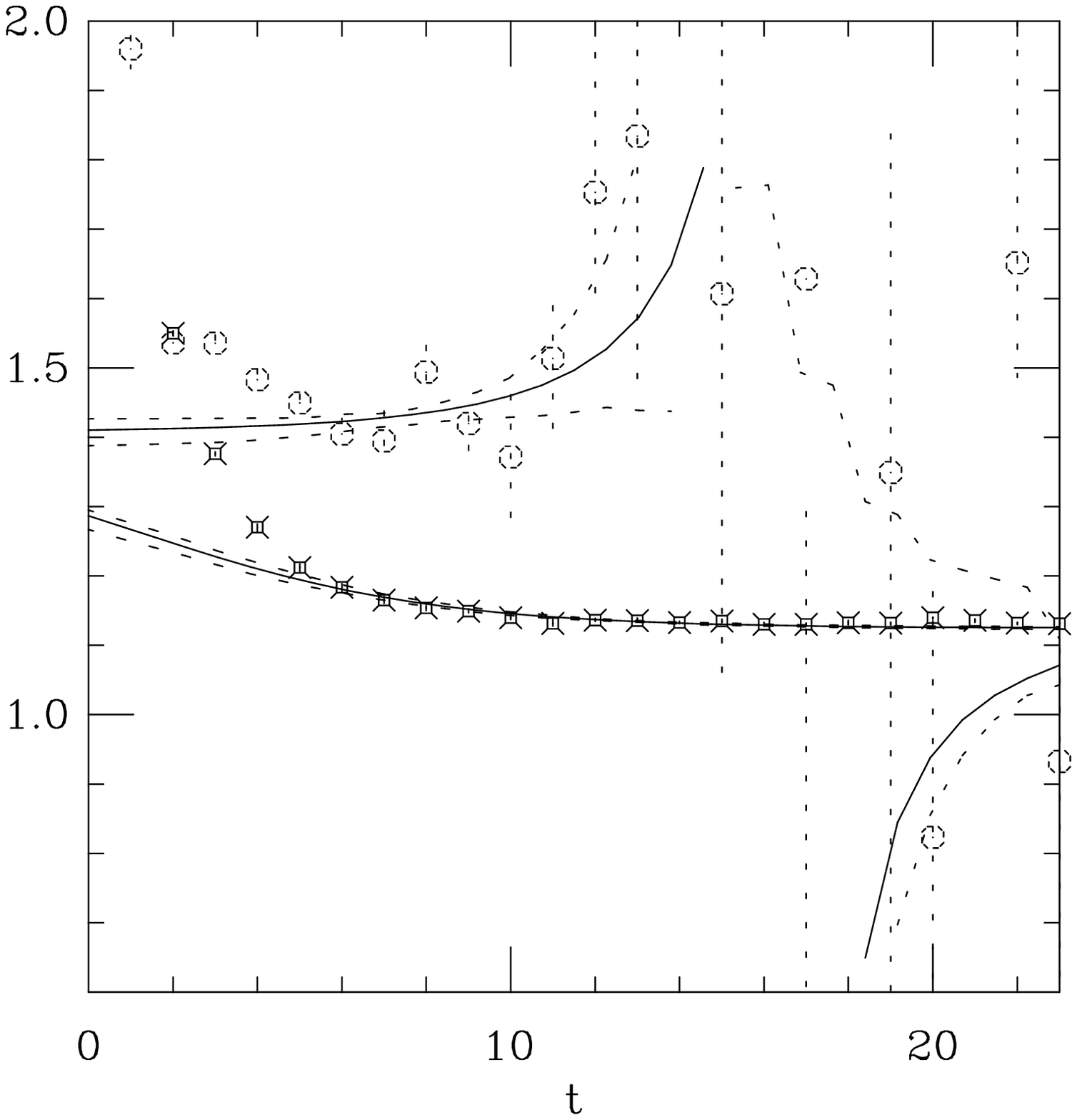}}
{FigHHDsmear}

\fltfig{Fit to the doubly $R_{10}$ source smeared with local sink
correlator (fancy squares) and the $R_{20}$ source smeared with local
sink correlator (circles) for the light-light meson.}
{\pspicture{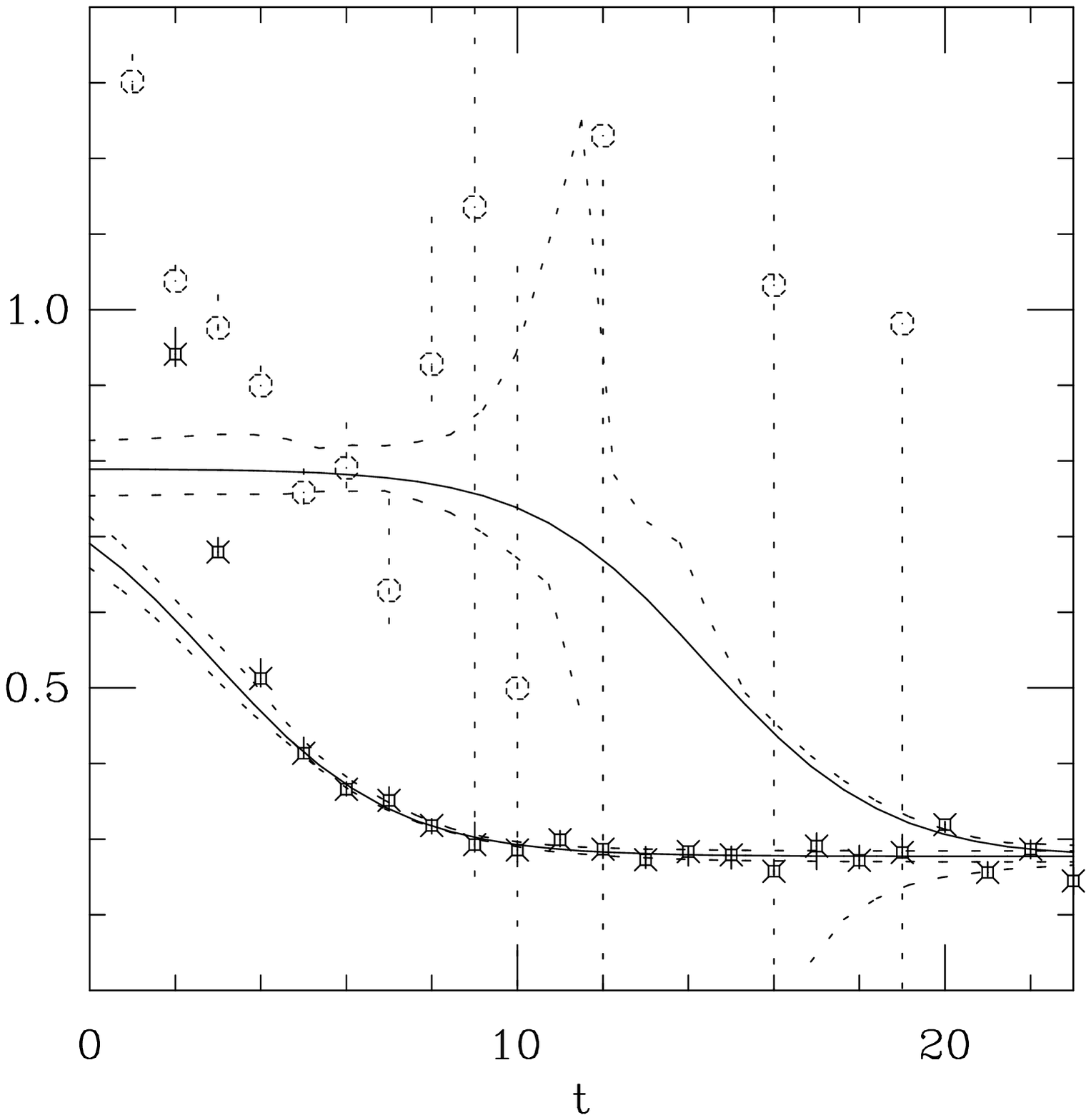}}
{FigLLDsmear}

\fltfig{Heavy-heavy fuzzed (r=6) and $R_{10}$ comparison. Both correlation
functions are sink smeared with local sources on the same configurations.
The lower statistical scatter our smearing technique
is thought to be in part due to including a non-zero contribution
from the local current.}
{\pspicture{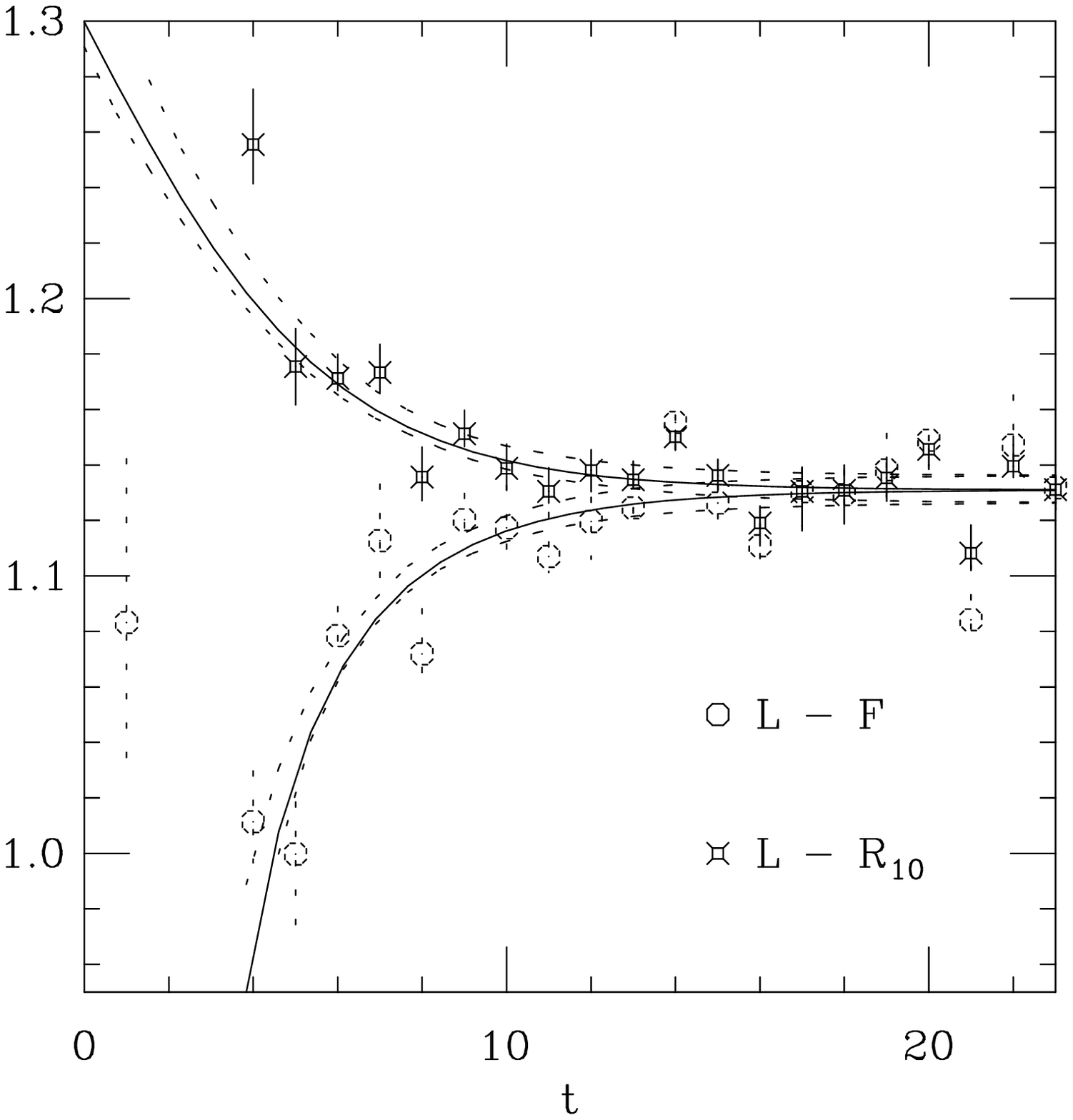}}
{FigHHfuzzComp}
\fltfig{Heavy-light fuzzed (r=6) and $R_{10}$ comparison.
Both correlation
functions are sink smeared with local sources on the same configurations.}
{\pspicture{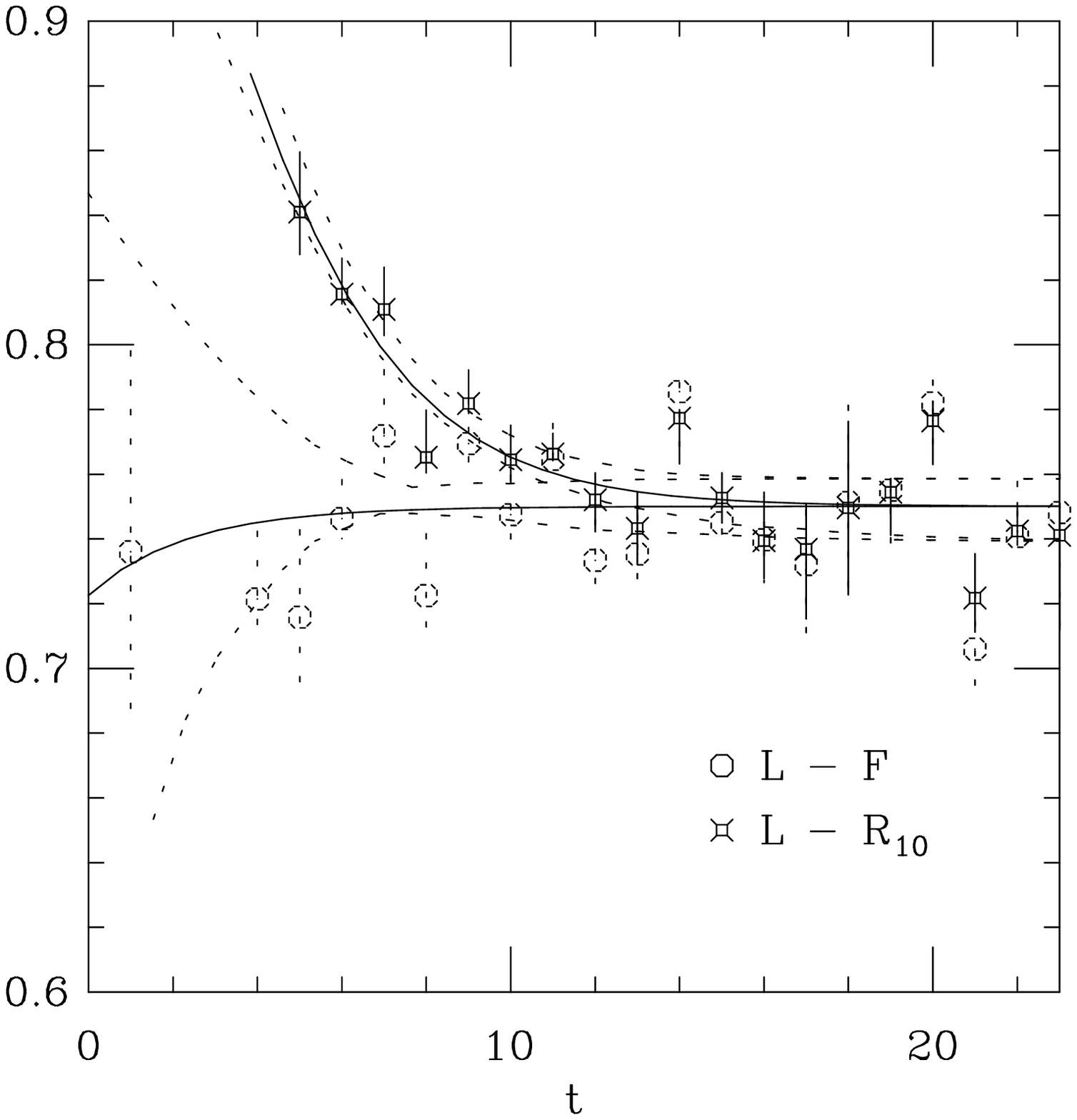}}
{FigHLfuzzComp}
\fltfig{Light-light fuzzed (r=8) and $R_{10}$ comparison.
Both correlation
functions are sink smeared with local sources on the same configurations.}
{\pspicture{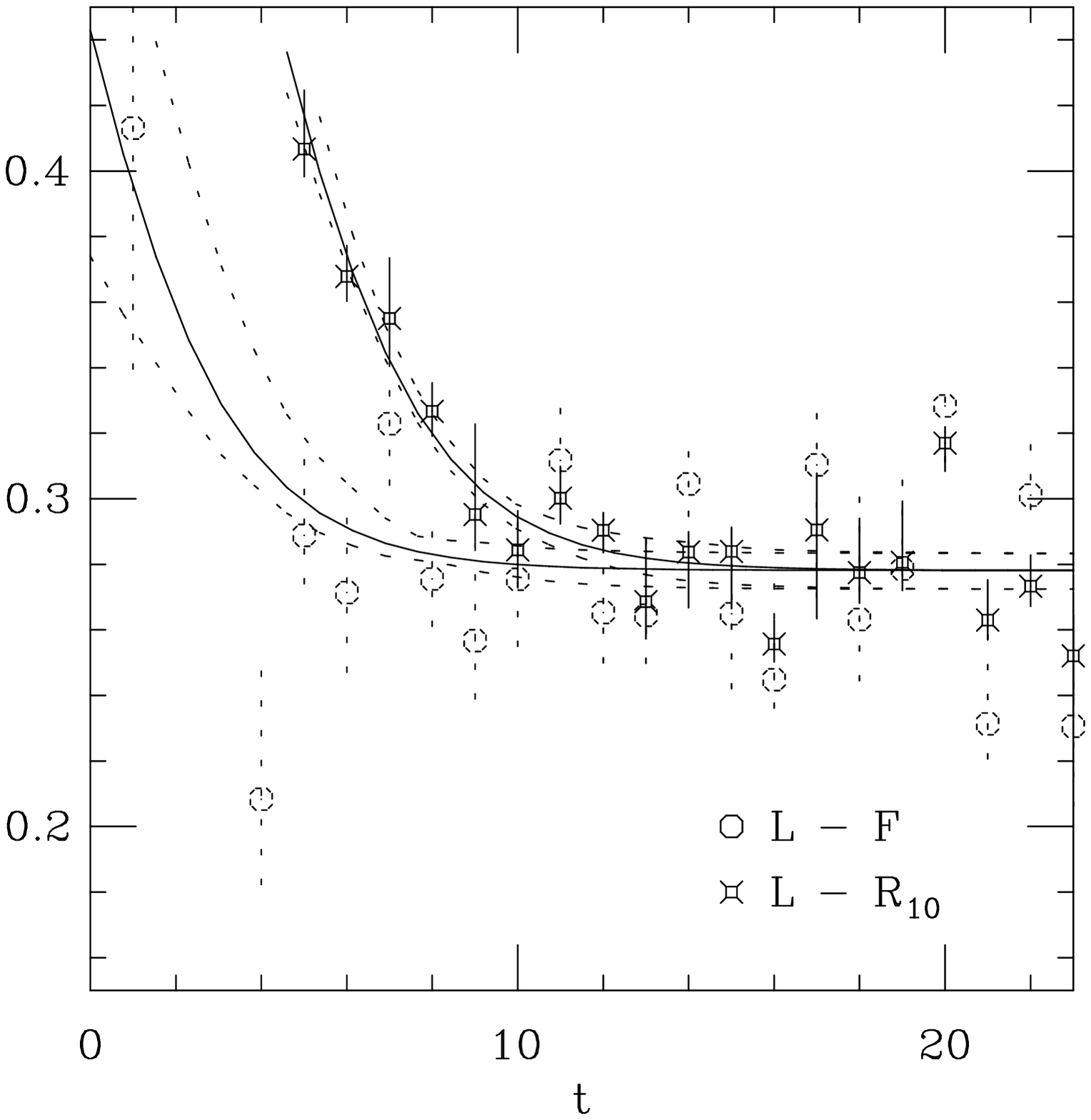}}
{FigLLfuzzComp}

\fltfig{Examples of approaches to fuzzed plateaux with high statistics. 
Opposing overlaps with the second and third excited states can
cause non-monotonic approaches. This is shown in the vector plot, where
the open circles represent the fuzzed source, local sink state.
The pseudoscalar plot shows a false plateau in the fuzzed-local
correlation function, since the fuzzed-fuzzed correlator forms
a plateau later we know that the early points in fuzzed-local
correlator plateau \emph{cannot} contain pure information about the
groundstate.}
{
\epsfaxhax
{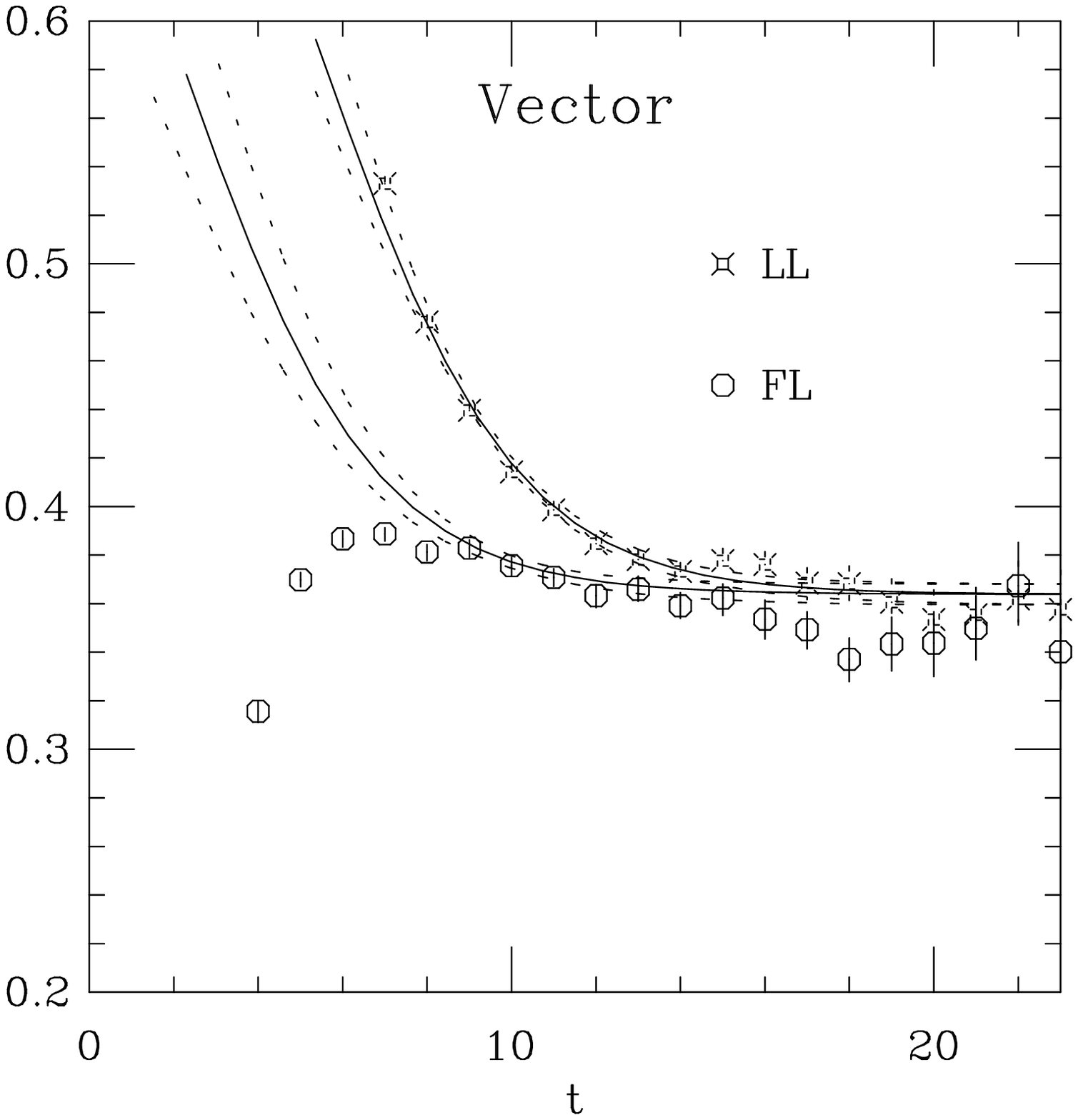}
{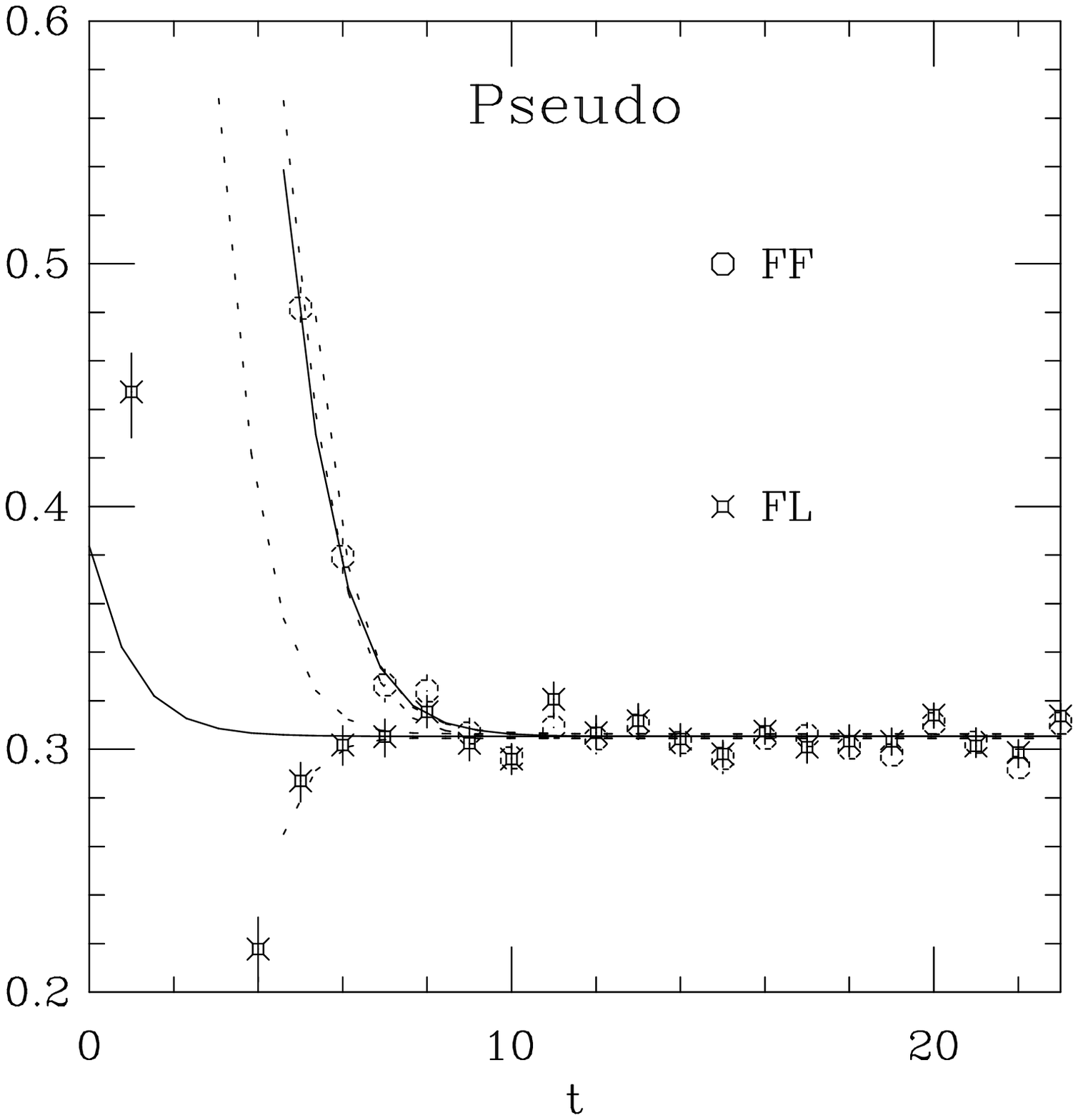}
}
{FigFuzzing}

\flttab{Smearing combinations}
{
\begin{tabular}{ccc|ccc}
$\kappa_1$& Source & Sink & $\kappa_2$ & Source & Sink \\
\hline
2300 & $L$      & $L$   & 2300 & $L$    & $L$   \\
2300 & $L$      & $F$   & 2300 & $L$    & $L$   \\
2300 & $R_{10}$& $L$    & 2300 & $L$    & $L$   \\
2300 & $R_{10}$& $L$    & 2300 & $R_{10}$ & $L$         \\
2300 & $R_{10}$ & $R_{10}$ & 2300 & $L$         & $L$   \\
2300 & $L$& $R_{20}$    & 2300 &  $L$   & $L$   \\
2300 & $R_{10}$& $R_{20}$& 2300 &  $R_{10}$     & $L$   \\
\hline
2300 & $L$      & $L$   & 3460 & $L$    & $L$   \\
2300 & $L$      & $F$   & 3460 & $L$    & $L$   \\
2300 & $R_{10}$         & $L$   & 3460 & $L$    & $L$   \\
2300 & $R_{10}$         & $R_{10}$      & 3460 & $L$    & $L$   \\
2300 & $L$      & $R_{10}$      & 3460 & $L$    & $L$   \\
2300 &  $L$     &  $R_{20}$     & 3460 & $L$    & $L$   \\
\hline
3460 & $L$      & $L$   & 3460 & $L$    & $L$   \\
3460 & $L$      & $F$   & 3460 & $L$    & $L$   \\
3460 & $L$      & $R_{20}$      & 3460 & $L$    & $L$   \\
3460 & $L$      & $R_{10}$      & 3460 & $L$    & $L$   \\
3460 & $L$      & $R_{10}$      & 3460 & $L$    & $R_{10}$      \\
\end{tabular}
}
{TabFeasSmearingComb}

\flttab{Heavy heavy double exponential fit stability; $r_0 = 2.6$, $T_{max}=20$}
{\begin{tabular}[h]{c|c|c|c|c}
tmin &$\chi^2/dof$& $\chi^2$	& Pseudo	& Pseudo$^\prime$\\
 \hline
2	& 0.66	& 21.	& 1.123(3)	& 1.45(2)\\
3	& 0.25	& 7.5	& 1.125(4)	& 1.42(1)\\
4	& 0.13	& 3.7	& 1.127(4)	& 1.41(2)\\
5	& 0.13	& 3.3	& 1.128(4)	& 1.40(3)\\
6	& 0.13	& 3.2	& 1.129(4)	& 1.40(3)\\
7	& 0.14	& 3.1	& 1.130(5)	& 1.40(4)\\
\end{tabular}
}
{TABHHStab}
\flttab{Heavy heavy triple exponential fit stability; $r_0 = 2.6$, $T_{max}=20$}
{\begin{tabular}[h]{c|c|c|c|c|c}
tmin &$\chi^2/dof$& $\chi^2$	& Pseudo (2300,2300)	& Pseudo$^\prime$	& Pseudo$^{\prime\prime}$\\
 \hline
2 	& 0.081	& 3.6	& 1.128(4)	& 1.39(3)	& 2.13(4)\\
3 	& 0.083	& 3.5	& 1.129(5)	& 1.39(3)	& 2.10(10)\\
4 	& 0.086	& 3.3	& 1.130(6)	& 1.39(4)	& 1.9(2)\\
5 	& 0.089	& 3.2	& 1.130(5)	& 1.40(7)	& 1.9(2)\\
6 	& 0.095	& 3.1	& 1.131(6)	& 1.39(8)	& 1.8(2)\\
\end{tabular}
}
{TABHHTripStab}
\flttab{Heavy light double exponential fit stability; $r_0 = 3.0$, $T_{max}=20$}
{\begin{tabular}[h]{c|c|c|c|c}
tmin &$\chi^2/dof$& $\chi^2$	& Pseudo& Pseudo$^\prime$\\
 \hline
2 & 0.44	& 14.	& 0.746(5)	& 1.18(2)\\
3 & 0.15	& 4.5	& 0.745(6)	& 1.13(2)\\
4 & 0.074	& 2.1	& 0.744(6)	& 1.10(2)\\
5 & 0.076	& 2.0	& 0.744(7)	& 1.09(3)\\
6 & 0.078	& 1.9	& 0.744(7)	& 1.08(4)\\
7 & 0.083	& 1.8	& 0.744(7)	& 1.07(6)\\
\end{tabular}
}
{TABHLStab}
\flttab{Light light double exponential fit stability; $r_0 = 3.1$, $T_{max}=18$}
{\begin{tabular}[h]{c|c|c|c|c|c}
tmin &$\chi^2/dof$& $\chi^2$	& Pseudo	& Pseudo$^\prime$\\
 \hline
2	& 0.18	& 5.0	& 0.280(7)	& 0.86(4)\\
3	& 0.085	& 2.2	& 0.279(7)	& 0.81(3)\\
4	& 0.079	& 1.9	& 0.278(7)	& 0.79(4)\\
5	& 0.082	& 1.8	& 0.278(7)	& 0.81(6)\\
\end{tabular}
}
{TABLLStab}

\flttab{Heavy heavy double exponential fit stability with Bohr radius}
{\begin{tabular}[h]{c|c|c|c|c|c|c}
Bohr & tmin  & $ \chi^2/dof $  & $ \chi^2 $      & Pseudo   & Pseudo $ ^\prime$ \\
\hline
2.60
&
4 	& 0.13	& 3.7	& 1.127(4)	& 1.41(2)\\
2.75
&
4 	& 0.083	& 2.3	& 1.128(4)	& 1.40(2)\\
2.80
&
4 	& 0.066	& 1.9	& 1.128(4)	& 1.40(2)\\
2.90
&
4 	& 0.045	& 1.2	& 1.128(4)	& 1.39(2)\\
3.00
&
4 	& 0.033	& 0.92	& 1.128(4)	& 1.39(2)\\
3.10
&
4 	& 0.026	& 0.74	& 1.128(4)	& 1.39(1)\\
3.25
&
4 	& 0.021	& 0.59	& 1.128(4)	& 1.39(1)\\
4.00
&
4 	& 0.016	& 0.44	& 1.128(3)	& 1.39(1)\\
5.50
&
4 	& 0.016	& 0.44	& 1.129(3)	& 1.40(6)\\
\end{tabular}
}
{TabHHBohrStab}

\flttab{Heavy light double exponential fit stability with Bohr radius}
{\begin{tabular}[h]{c|c|c|c|c|c}
Bohr & tmin  & $ \chi^2/dof $  & $ \chi^2 $      & Pseudo  & Pseudo $ ^\prime$\\
\hline
2.60
&
4 	& 0.059	& 1.6	& 0.745(6)	& 1.13(3)\\
2.75
&
4 	& 0.10	& 2.8	& 0.745(6)	& 1.12(2)\\
2.80
&
4 	& 0.12	& 3.2	& 0.745(6)	& 1.12(2)\\
2.90
&
4 	& 0.12	& 3.3	& 0.744(6)	& 1.11(2)\\
3.00
&
4 	& 0.074	& 2.1	& 0.744(6)	& 1.10(2)\\
3.10
&
4 	& 0.045	& 1.2	& 0.744(6)	& 1.10(2)\\
3.25
&
4 	& 0.026	& 0.72	& 0.744(6)	& 1.09(2)\\
4.00
&
4 	& 0.013	& 0.35	& 0.744(6)	& 1.09(3)\\
5.50
&
4 	& 0.014	& 0.38	& 0.745(6)	& 1.10(5)\\
\end{tabular}
}
{TabHLBohrStab}

\flttab{Light light double exponential fit stability with Bohr radius}
{\begin{tabular}[h]{c|c|c|c|c|c}
Bohr & tmin & $ \chi^2/dof $  & $ \chi^2 $      & Pseudo   & Pseudo $ ^\prime$ \\
\hline
2.60
&
4 	& 0.0053	& 0.13	& 0.278(7)	& 0.79(4)\\
2.75
&
4 	& 0.010	& 0.24	& 0.278(7)	& 0.78(4)\\
2.80
&
4 	& 0.013	& 0.32	& 0.278(7)	& 0.78(4)\\
2.90
&
4 	& 0.029	& 0.70	& 0.278(7)	& 0.78(4)\\
3.00
&
4 	& 0.077	& 1.9	& 0.278(7)	& 0.78(3)\\
3.10
&
4 	& 0.079	& 1.9	& 0.278(7)	& 0.79(4)\\
3.25
&
4 	& 0.023	& 0.55	& 0.278(7)	& 0.79(4)\\
4.00
&
4 	& 0.0013	& 0.031	& 0.277(6)	& 0.77(7)\\
5.50
&
4 	& 0.00072	& 0.017	& 0.277(6)	& 0.8(1)\\
\end{tabular}
}
{TabLLBohrStab}

\flttab{Optimal Bohr radii}
{
\begin{tabular}{ccc}
System& Fitted Value & Choice\\
\hline
2300-2300 & 2.62(2) & 2.6\\
2300-3460 & 2.98(2) & 3.0\\
3460-3460 & 3.09(1) & 3.1\\
\end{tabular}
}
{TabOptimalRadii}

\flttab{Comparison of 2S-1S splittings with experiment using string tension scale}
{
\begin{tabular}{ccc}
System &  Lattice (MeV)&  Experimental (MeV)\\
\hline
Light Pseudoscalar & 1370(50) & 1170(100) ($\pi$)\\
Light Vector & 1250(100)& 700(25) ($\rho$)\\
Heavy-Light Pseudoscalar & 970(60) & - \\
Heavy-Heavy Pseudoscalar & 710(60) & 590 ($J/\psi$)\\
\end{tabular}
}
{TabRadExpComp}

\end{document}